\documentclass[pdflatex,sn-mathphys-num]{sn-jnl}

\usepackage{graphicx}%
\usepackage{multirow}%
\usepackage{amsmath,amssymb,amsfonts}%
\usepackage{amsthm}%
\usepackage{mathrsfs}%
\usepackage[title]{appendix}%
\usepackage{xcolor}%
\usepackage{textcomp}%
\usepackage{manyfoot}%
\usepackage{booktabs}%
\usepackage{algorithm}%
\usepackage{algorithmicx}%
\usepackage{algpseudocode}%
\usepackage{listings}%
\usepackage{amsmath}
\usepackage{array}         
\usepackage{booktabs}       
\usepackage{multirow}      
\usepackage{enumitem}

\theoremstyle{thmstyleone}%
\theoremstyle{thmstyletwo}%

\theoremstyle{thmstylethree}%

\begin{document}

\title[Article Title]{Recent Advances in Malware Detection: Graph Learning and Explainability}


\author[1]{\fnm{Hossein} \sur{Shokouhinejad}}

\author[1]{\fnm{Roozbeh} \sur{Razavi-Far}}

\author[1]{\fnm{Hesamodin} \sur{Mohammadian}}

\author[1]{\fnm{Mahdi} \sur{Rabbani}}

\author[1]{\fnm{Samuel} \sur{Ansong}}

\author[1]{\fnm{Griffin} \sur{Higgins}}

\author[1]{\fnm{Ali A} \sur{Ghorbani}}

\affil*[1]{\orgdiv{University of New Brunswick}, \orgname{Faculty
 of Computer Science}, \orgaddress{\street{3 Bailey Dr.}, \city{Fredericton}, \postcode{E3B5A3}, \state{New Brunswick}, \country{Canada}}}




\abstract{The rapid evolution of malware has necessitated the development of sophisticated detection methods that go beyond traditional signature-based approaches. Graph learning techniques have emerged as powerful tools for modeling and analyzing the complex relationships inherent in malware behavior, leveraging advancements in Graph Neural Networks (GNNs) and related methods. This survey provides a comprehensive exploration of recent advances in malware detection, focusing on the interplay between graph learning and explainability. It begins by reviewing malware analysis techniques and datasets, emphasizing their foundational role in understanding malware behavior and supporting detection strategies. The survey then discusses feature engineering, graph reduction, and graph embedding methods, highlighting their significance in transforming raw data into actionable insights, while ensuring scalability and efficiency. Furthermore, this survey focuses on explainability techniques and their applications in malware detection, ensuring transparency and trustworthiness. By integrating these components, this survey demonstrates how graph learning and explainability contribute to building robust, interpretable, and scalable malware detection systems. Future research directions are outlined to address existing challenges and unlock new opportunities in this critical area of cybersecurity.}




\maketitle

\section{Introduction}
Malware or malicious software is purposefully designed to damage digital devices and compromise their functionality. The growing sophistication of malware reflects its increasing prevalence and impact. Between 2019 and 2024, the malware analysis market is expected to expand significantly, from 3 billion USD to 11.7 billion USD, with a compound annual growth rate (CAGR) of 31 percent \cite{marketsandmarkets}. According to the AV-Test Institute, over 360,000 new malware samples are generated daily, equating to roughly 4.2 new malware every second \cite{avtest2019}. Over the past five years, malware has been increasing by 100 million annually, necessitating advanced mitigation strategies to address this surge. While earlier malware often featured simplistic code and could be detected with basic tools, modern malware has become increasingly complex. Sophisticated malware now often evades traditional defenses such as firewalls and antivirus software, remains undetected in systems for extended periods, and propagates through networks with devastating impact.

\begin{figure}[h]
    \centering
    \includegraphics[width=0.5\linewidth]{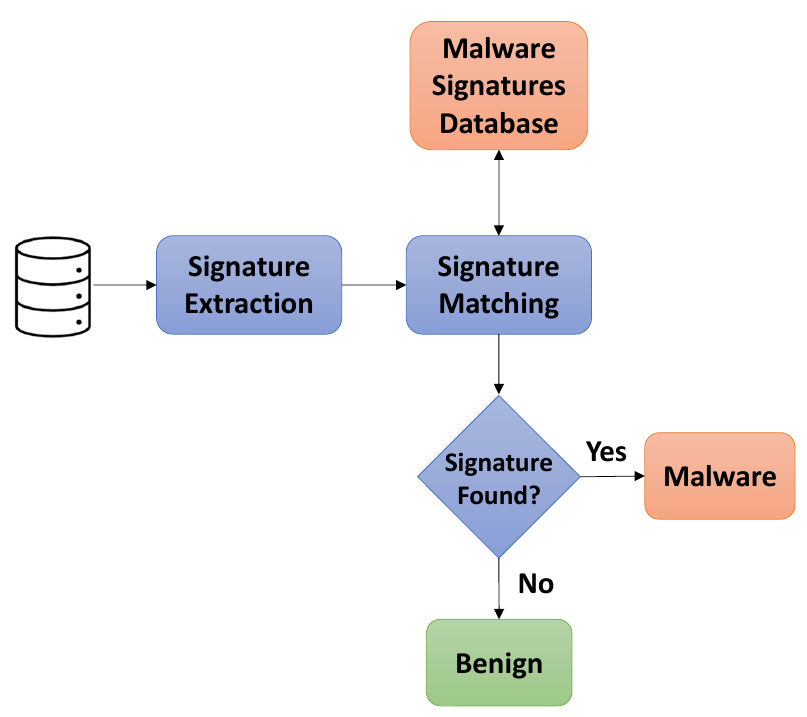}
    \caption{General diagram for signature-based malware detection systems.}
    \label{fig:second_image}
\end{figure}

One of the foundational methods in malware detection is signature-based detection, which relies on predefined patterns to identify malicious software. As illustrated in Figure \ref{fig:second_image}, this approach remains a cornerstone in cybersecurity, despite its limitations. Efforts to enhance signature-based detection have explored innovative techniques, such as n-grams and opcode sequence analysis \cite{Santos2009,Santos2013}. These methods improve the identification of unknown malware by analyzing file signatures and executable representations. Research has also extended signature-based detection to application-specific contexts. For instance, Ngamwitroj et al. \cite{Ngamwitroj2018} successfully adapted these techniques for the Android platform, utilizing permission and broadcast-receiver data to achieve notable accuracy. However, as malware becomes more sophisticated, the limitations of traditional signature-based methods become evident. Comprehensive reviews, such as \cite{8949524}, emphasize these challenges, highlighting the critical need for continued evolution in detection techniques to safeguard digital environments.

\begin{figure}[h]
    \centering
    \includegraphics[width=0.5\linewidth]{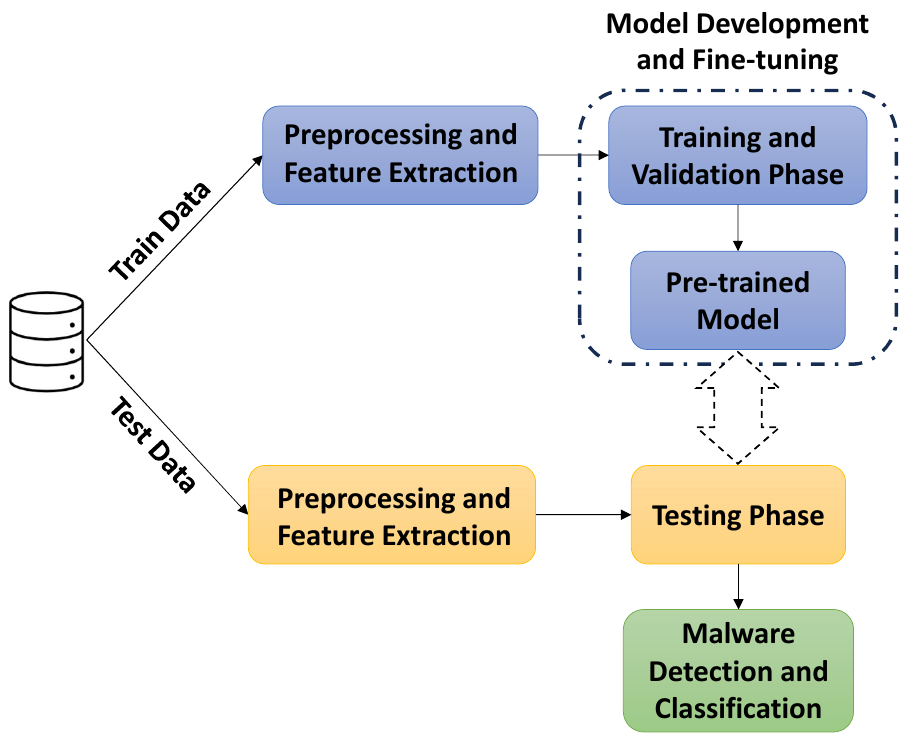}
    \caption{General diagram for ML-based malware detection systems.}
    \label{fig:first_image}
\end{figure}

Machine Learning (ML) approaches, including both classical machine learning and deep learning, have revolutionized malware detection by overcoming the limitations of traditional signature-based methods. Figure \ref{fig:first_image} illustrates the general diagram of ML-based malware detection systems. These techniques adapt to evolving threats, improve detection rates, reduce false positives, and addressing challenges like zero-day malware and obfuscation. Classical ML methods such as Decision Trees, Support Vector Machines (SVM), k-Nearest Neighbors (k-NN), and Random Forests excel in feature-based classification by analyzing the statistical properties of software to identify potential malware. For instance, Random Forests mitigate overfitting through ensemble learning, making them highly effective for malware classification \cite{zhang2016robust}, while SVMs demonstrate high accuracy in binary classification problems by identifying optimal hyperplanes in high-dimensional feature spaces \cite{gandotra2014malware}. Similarly, k-NN, though computationally intensive, achieves remarkable accuracy by leveraging proximity-based classification \cite{kolter2006learning}.
Deep learning techniques, including Convolutional Neural Networks (CNNs), Recurrent Neural Networks (RNNs), and Graph Neural Networks (GNNs), further advance malware detection by automating feature extraction and learning directly from raw data. CNNs excel in treating binary files as images, automatically extracting intricate features \cite{nataraj2011malware}, while RNNs, particularly Long Short-Term Memory (LSTM) networks, are highly effective at analyzing sequential data, enabling dynamic behavior analysis of malware over time \cite{pascanu2015malware}.
Despite the success of classical and deep learning-based malware detection techniques, traditional feature representations often fail to capture the complicated structural and relational dependencies within malware behavior. Many existing methods rely on individual attributes or sequential data representations, which do not fully reflect the interconnected nature of malicious activities. The use of graph data can address this limitation by encoding relationships between system calls, control flow structures, and API interactions, enabling a more comprehensive analysis of malware behavior through graph learning techniques. By leveraging graph structures, malware detection models can identify shared subgraph patterns, measure similarity between malware families, and uncover hidden relationships that conventional methods might miss. GNNs further enhance this approach by learning hierarchical, neighborhood-based representations of malware execution, capturing both local and global patterns. Unlike CNNs or RNNs, which operate on grid-like or sequential data, GNNs model malware behavior as graphs, enabling scalable, explainable, and adaptive detection strategies against evolving cyber threats.

Handling large and complex graphs is a significant challenge in graph-based malware detection. The high computational cost and memory requirements associated with processing large-scale graphs often hinder the scalability of ML models, particularly for malware analysis, where graphs can represent intricate relationships such as control flow or Application Programming Interface (API) call dependencies. To address this, graph reduction techniques have emerged as effective solutions, simplifying graph structures, while preserving critical information. These techniques, such as node and edge pruning reduce graph size, enabling faster computations and improved scalability without compromising the structural or topological properties essential for detection accuracy \cite{CoreWalk}. By leveraging reduced graphs, GNNs efficiently learn and infer malware patterns, utilizing the structural relationships encoded in graph data. GNNs excel at capturing both local and global dependencies within reduced graphs, enhancing their applicability in malware detection tasks, from identifying malicious subgraphs to explaining decision. This synergy between graph reduction and GNNs ensures scalable and robust detection systems capable of adapting to evolving malware threats.

As graph reduction and GNNs address the challenges of scalability and efficiency in malware detection, explainability emerges as a complementary and vital component, ensuring that these advanced methods remain interpretable and actionable. Explainability is essential in artificial intelligence (AI), particularly in cybersecurity, where understanding the reasoning behind model predictions builds trust, supports regulatory compliance, and aids in identifying biases or errors \cite{bose2020explaining}. In GNNs, explainability techniques seek to uncover the underlying factors driving model predictions by identifying critical graph elements—such as nodes, edges, or substructures—that influence outcomes \cite{amara2024graphframexsystematicevaluationexplainability}. This interpretability is especially important given the complexity of graph-based models.
In malware detection, explainability enhances both detection and response efforts by revealing patterns and features indicative of malicious behavior. For instance, it can identify anomalous subgraphs, irregular control flows, or suspicious API call dependencies that are key to classifying malware \cite{CFG_4}. These insights empower security analysts to validate model outputs, understand attack vectors, and design proper mitigation strategies. As malware evolves, explainability ensures that AI systems not only maintain high detection accuracy but also provide actionable and transparent insights, making them indispensable tools in the fight against cyber threats.
\begin{figure*}[h]
    \centering
    \includegraphics[width=\linewidth]{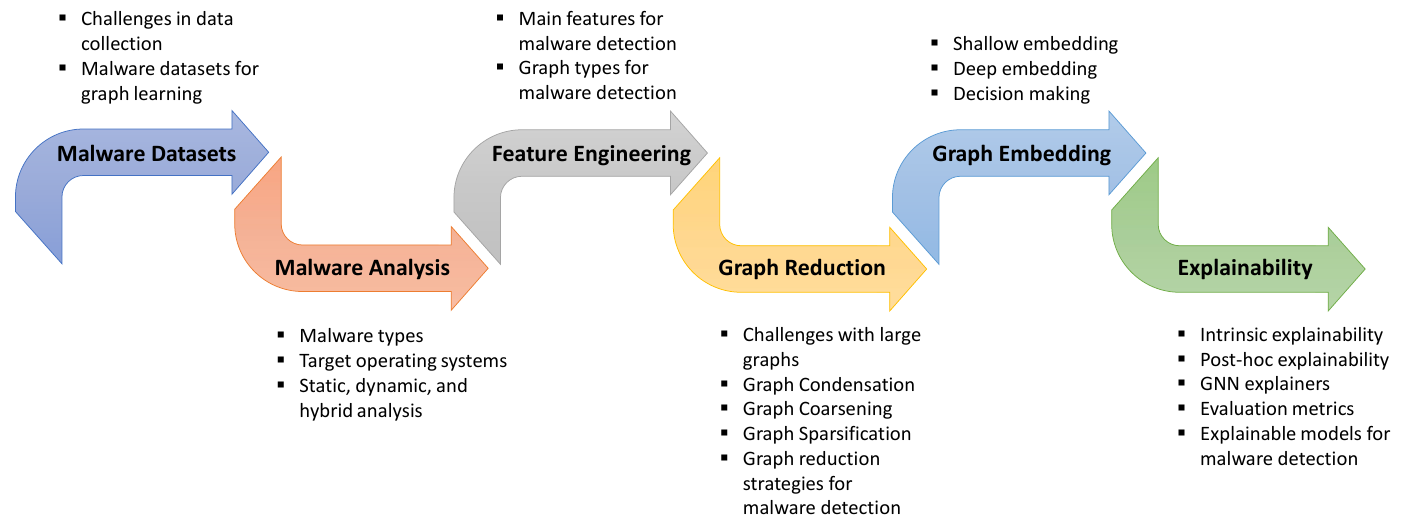}
    \caption{Roadmap for recent advances in graph-based malware detection, showcasing the interconnected sections of the survey, from foundational malware analysis to explainable malware detection systems.}
    \label{fig:survey_pipeline}
\end{figure*}

To address the dynamic challenges of malware detection, this survey provides a comprehensive exploration of recent advances in graph learning and explainability. Organized into eight interconnected sections, the survey outlines the critical components of the malware detection pipeline, illustrating their roles and interplay in building robust detection systems.
Section 2 highlights the importance of malware datasets, addressing challenges in data collection, benchmarking, and the specific needs of graph learning-based approaches, while identifying prospective datasets to advance malware detection research.
Section 3 examines malware analysis, exploring static, dynamic, and hybrid analysis techniques, as well as examining various malware types and their target operating systems. This comprehensive analysis provides a foundation for understanding malware behavior and developing effective detection strategies.
Section 4 focuses on feature engineering, where raw data is processed into meaningful representations. This step bridges the gap between data collection and model application, ensuring that traditional ML and graph-based methods are supplied with relevant, informative features.
Section 5 explores graph reduction techniques in depth, providing a comprehensive overview of current methods and evaluating their suitability for graph-based malware detection. These techniques address the computational challenges posed by large and complex graphs by reducing their size while preserving critical topological information, enabling scalable and efficient analysis.
Section 6 provides an extensive review of current graph embedding techniques. It explores the strengths and limitations of these methods, evaluating their suitability for malware detection. These embeddings transform graph data into representations that capture both structural and relational properties, facilitating malware classification.
Section 7 emphasizes the importance of explainability for malware detection, initially explores explainable artificial intelligence (XAI) methods in general, and, then, focuses on GNN explainers. Next, it explores existing works applying explainability techniques to malware detection.
Finally, Section 8 presents the conclusion, summarizing the insights gained from this survey and discussing future research directions.

This survey not only reviews state-of-the-art techniques but also demonstrates their interconnection, showcasing how each component contributes to create scalable, interpretable, and effective malware detection systems. The overall pipeline is visualized in Figure \ref{fig:survey_pipeline}, which illustrates the logical flow of the survey's structure.

\section{Malware Datasets}
As ML methods have become more integral to identify malware, the need for collection of comprehensive, accurately labeled, and diverse datasets has grown exponentially. These datasets not only fuel the development of more robust and resilient detection models but also enable researchers to stay ahead in the arms race against cyber threats. However, the process of data collection in the realm of malware presents a unique set of challenges, ranging from the sheer diversity and adaptability of malware to legal and ethical considerations. Furthermore, benchmark datasets are essential for evaluating the performance of ML models and fostering reproducibility in research, which introduces additional layers of complexity.
Beyond these challenges, the emergence of graph learning-based techniques in malware detection has underscored the need for datasets that include structural features, such as program execution flow, which are crucial for these models. Many commonly used datasets fail to provide access to raw binaries or graph-relevant features, limiting their utility in this domain.

\subsection{Challenges in Data Collection}
An essential first step in the development of ML models involves the collection of data that accurately reflects the distribution of binaries observed in real-world circumstances. The improvement of ML systems' performance is commonly achieved by raising the quality and amount of labeled data, as supported by known methods \cite{Domingos2012,Halevy2009}. Nevertheless, the scope of potential binary behaviors is boundless, rendering it impractical to randomly select from the vast array of preexisting binaries. Therefore, it presents a difficulty to determine the complete extent of coverage that given datasets possess throughout the entire range of binary possibilities. The presence of malware in the domain presents further challenges for gathering data, making it impossible to use conventional methods such as using numerous annotators for each file and evaluating their consensus \cite{Geiger2020}.

Surprisingly, malware data is more easily accessible compared to innocuous data. The accessibility of malware samples can be attributed to the presence of platforms that aggregate submissions from volunteers \cite{Roberts2011,Quist2009} and the implementation of honeypots by researchers \cite{Baecher2006}. A honeypot is an internet-connected configuration specifically created to attract and catch malware by intentionally revealing weaknesses and avoiding typical security procedures. Nevertheless, the data obtained from honeypots and malware submission sites remains susceptible to certain limitations. The data obtained from honeypots has an inherent bias towards the malware that can be attracted by the particular configuration of the honeypot. For example, some malware may only become active when it comes into contact with specific apps, necessitating the honeypot to imitate those circumstances \cite{Zhuge2007}. In addition, certain malware has the ability to detect and evade honeypots, hence distorting the acquired data \cite{Krawetz2004}. These problems have an impact not just on the data stored in honeypots but also on the reliability of bigger malware databases, particularly if the contributions originate from honeypot operators. Moreover, these repositories may demonstrate a prejudice as a result of the persons' willingness and capability to provide malware samples.

Conversely, the acquisition of benign data presents much more formidable obstacles. In contrast to malware, benign applications do not aim to spread throughout the internet, which complicates their collecting process. Thus far, there has been a dearth of scholarly investigations pertaining to the diversity and procurement of benign samples. The common approach has been to collect data from the host operating system, which may introduce bias if proprietary features are shared. Studies have shown that models can distinguish binaries based solely on their affiliation with the operating system, leading to the misclassification of all operating system binaries as malicious \cite{Krawetz2004,Seymour2016}. This methodology has resulted in widespread biases, as evidenced by the heavy reliance on Windows binaries in various research studies. Nevertheless, it is worth noting that corporate entities engaged in the development of antivirus solutions deviate from the norm due to their exclusive access to proprietary datasets.

\subsection{Prospective Malware Datasets for Graph Learning}

While many commonly used malware detection datasets exist, they often follow a structure that provides open access to extracted features (derived from both benign and malicious raw binaries) and request based access to raw binaries, typically malicious, if at all. Generally, this structure has the benefit of addressing legal and copyright issues via controlling access to the binaries while allowing open dissemination of the extracted features needed to train and test traditional ML models. Unfortunately, this presents a large challenge for graph learning based techniques that rely on structural features often related to program execution flow. As these features are not included in the extracted features they must be directly generated from raw binaries. However, since access to raw binaries, especially benign, is often restricted this task becomes even more challenging.

In Table \ref{tab:mal-dataset}, we include several prospective datasets specifically can be used for graph learning and detecting malware. Thus, we omit several common datasets that do not include raw binaries or features suitable for graph learning, such as \cite{anderson2018ember}, or are outdated and no longer actively served, such as \cite{hutchison_adam_2013}. Furthermore, where feasible, we provide the estimated number of usable available benign and malicious binaries as well as their access level to facilitate graph learning for malware detection.

\begin{table*}[h]
\centering
\caption{Malware datasets with binary samples (D: Dataset, R: Repository, M: Marketplace, A: Android, W: Windows, L: Labeled, U: Unlabeled, O: Ongoing).}
\label{tab:mal-dataset}
\resizebox{\textwidth}{!}{%
\begin{tabular}{lcccccc}
\toprule
Dataset                                 & Type & \#Benign     & \#Malicious & Sample Type & Label & Binary Access \\ \midrule
PMML \cite{practicalsecurity2021pe}     & D & 86,812                           & 114,737                         & W & L& Open                              \\
DikeDataset \cite{dikedataset}          & D & 1,082                            & 10,841                          & W & L& Open                              \\
BODMAS \cite{bodmas}                    & D & Closed                           & 77,142                          & W & L & Request                           \\
CCCS-CIC-AndMal-2020 \cite{cic-cccs-andmal-2020-1}, \cite{cic-cccs-andmal-2020-2}                   & D & 200,000                          & 200,000                         & A & L & Open                  \\
CICMalDroid2020 \cite{cicmaldroid}      & D & 0                                & 17,341                          & A & L & Open                              \\
SOREL-20M \cite{harang2020sorel20m}     & D & Closed                           & 9,919,251                       & W & L & Open                              \\
CIC-AndMal2017 \cite{cicandmal}         & D & 4,354                            & 6,500                           & A & L & Open                              \\
Derbin \cite{arp2014drebin}             & D & Closed                           & 5,560                           & A & L & Open                              \\
AndroZoo \cite{AndroZoo}                & R & 24,870,486                      & 24,870,486                    & A & O & Request                           \\
VirusShare \cite{Roberts2011}           & R & 0                                & 93,049,515                      & A & L & Request                           \\
Chocolatey \cite{Chocolatey}            & M & 10,363 (packages)               & 10,363 (packages)              & A & U & Open                          \\
Google Play \cite{GooglePlay}           & M & 1.5-3M                           & 1.5-3M                        & A & U & Open                           \\
\bottomrule
\end{tabular}
}
\end{table*}

\section{Malware Analysis}
\begin{figure}[h]
    \centering
    \includegraphics[width=0.5\linewidth]{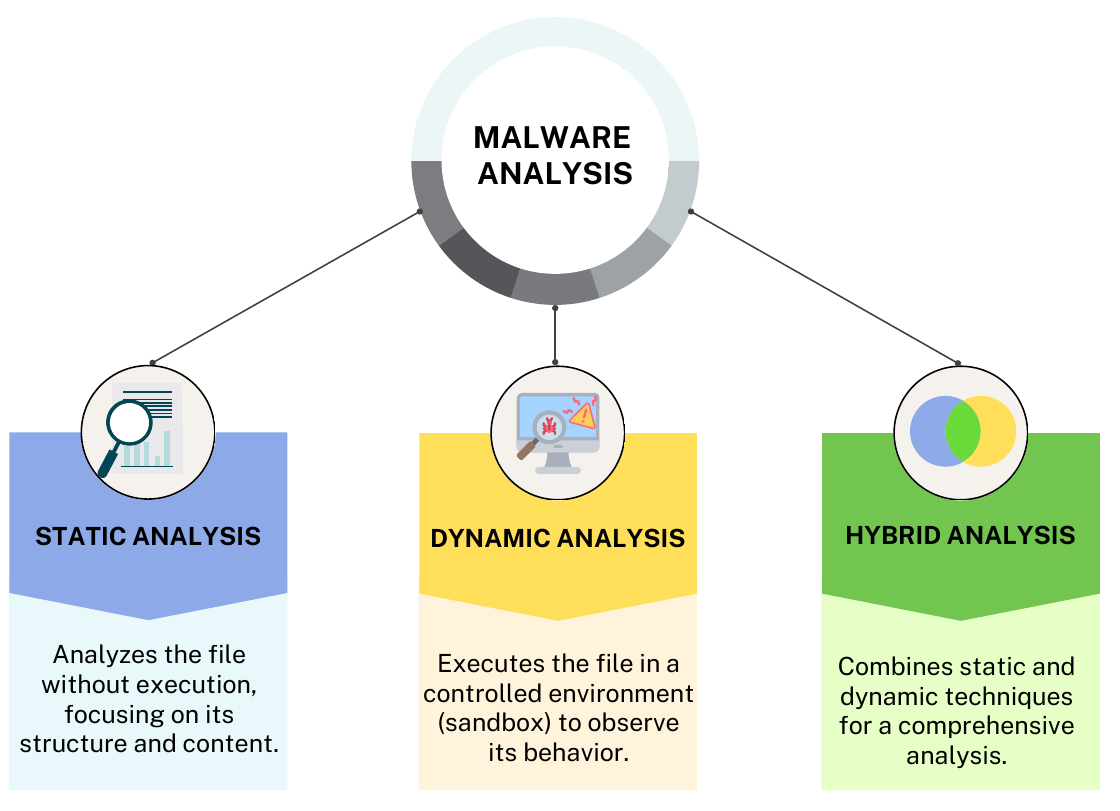}
    \caption{Main malware analysis approaches.}
    \label{Malware analysis}
\end{figure}

 Extensive research has been dedicated to analyzing malware from different perspectives of file or program properties, broadly categorized into three main groups: static analysis, dynamic analysis, and hybrid analysis, as illustrated in Figure \ref{Malware analysis}. This categorization reflects the multifaceted nature of file or program analysis, facilitating a comprehensive approach to feature extraction. The pie chart in Figure \ref{Pie Chart 2}, presents the percentage of the analysis methods adopted by researchers. The statistical data in this figure are derived from our extensive and systematic literature survey examining the research landscape on malware detection from 2015 to 2024. This was achieved using search terms like ``malware detection'' and ``malware classification''. We gathered about 500 pertinent research articles from two sources: IEEE Xplore and ScienceDirect.

\subsection{Static Analysis}
Static features are derived directly from the samples themselves without the need to execute them. Once the file has been unpacked and decrypted, it is possible to extract features used in static analysis. Static analysis can be performed at three levels: binary level, metadata level, and code level. At the binary level, features are extracted by inspecting the bytes. Features at the metadata level are extracted by analyzing details about the file format, such as information found in the PE header. To thoroughly analyze the file at the code level, it is necessary to convert the file into assembly code. Disassembly tools, such as IDA Pro \cite{hexrays}, are commonly used for this purpose. From the assembly code, features such as the execution flow can be extracted.
\subsection{Dynamic Analysis}
Dynamic analysis for the files involves executing the files in a controlled environment to observe their behavior. This method provides insights into the file's actual impact and capabilities, which may not be evident through the static analysis. During the dynamic analysis, the file is typically executed in a sandbox or virtual machine to prevent it from affecting the host system. Analysts monitor various aspects of the file's behavior, such as API and system calls, file system modifications, network activity, and interactions with system processes. The dynamic analysis is particularly useful for detecting polymorphic and packed malware that can evade those detection techniques based on the static analysis.
\subsection{Hybrid Analysis}
Hybrid analysis for malware detection combines the strengths of both the static and dynamic analyses. Static analysis, while cost-effective, can struggle with high performance due to techniques like obfuscation and packing. On the other hand, dynamic analysis is adept at detecting malware variants and new families, as it is resilient to low-level obfuscation. However, it can be costly and not scalable due to its limited coverage.  The hybrid analysis aims to overcome these limitations by integrating both approaches. For instance, a packed malware can first undergo the dynamic analysis to extract its hidden-code bodies by comparing its runtime execution with its static code model. Once these hidden-code bodies are uncovered, a static analyzer can then continue the analysis, combining the benefits of both approaches for a more effective malware detection process.
\begin{figure}[h]
    \centering
    \includegraphics[width=0.35\linewidth]{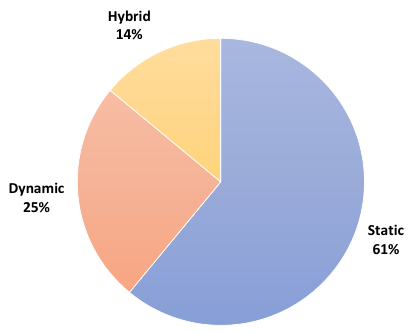}
    \caption{The percentage of malware analysis approaches derived from 500 research works on malware detection, collected from 2015 to 2024.}
    \label{Pie Chart 2}
\end{figure}

\subsection{Malware Types}
Malware refers to any software intentionally designed to cause damage to a computer, server, client, or computer network. Historically, malware has evolved significantly from simple viruses and worms to complex, multi-faceted threats that can adapt and evade traditional security measures. Early malware primarily focused on causing disruption or damage to systems, often spreading through floppy disks and network connections. Over time, the motivation behind malware shifted from mere disruption to financial gain, espionage, and information theft, leading to the development of more sophisticated forms such as ransomware, spyware, and advanced persistent threats (APTs) \cite{Razak2016}.

The increasing complexity of computing systems and the ubiquity of internet access have provided fertile ground for the evolution of malware. Modern malware often incorporates advanced techniques like polymorphism and obfuscation to avoid detection by traditional antivirus programs. ML and AI are now being employed to both development of more sophisticated malware and to create robust detection and mitigation strategies. Researchers have continually adapted to these changes, developing new methods to analyze and classify malware using dynamic and/or static analysis techniques \cite{Chandy2022, Tooth2019}. The most common types of malware are listed below, categorized by their unique characteristics and behavior.

\begin{itemize}[leftmargin=*]
  \item \textbf{Viruses}: Malware that attaches itself to clean files and spreads throughout a computer system, often damaging files and software \cite{Razak2016}.
  \item \textbf{Worms}: Self-replicating malware that spreads across networks without needing to attach itself to an existing file \cite{Bavishi2017}.
  \item \textbf{Trojans}: Malware disguised as legitimate software that tricks users into loading and executing it on their systems, often creating backdoors for unauthorized access \cite{Pirscoveanu2015}.
  \item \textbf{Spyware}: Malware designed to spy on a user's activities without their knowledge, often collecting personal information such as passwords and financial details \cite{Chandy2022}.
  \item \textbf{Ransomware}: Malware that encrypts a user's files and demands a ransom payment to restore access, causing significant disruption and financial loss \cite{Razak2016}.
  \item \textbf{Rootkits}: Malware that provides privileged access to a computer, while hiding its presence and the presence of other malicious software \cite{rootKits}.
  \item \textbf{Adware}: Malware that automatically delivers advertisements, often bundled with free software and can collect data on user behavior \cite{Gandotra2014}.
  \item \textbf{Botnets}: Networks of infected computers controlled by an attacker, often used to launch large-scale attacks such as DDoS (Distributed Denial of Service) \cite{Chandy2022}.
\end{itemize}

\subsection{Target Operating Systems}
\begin{figure}[h]
    \centering
    \includegraphics[width=0.8\linewidth]{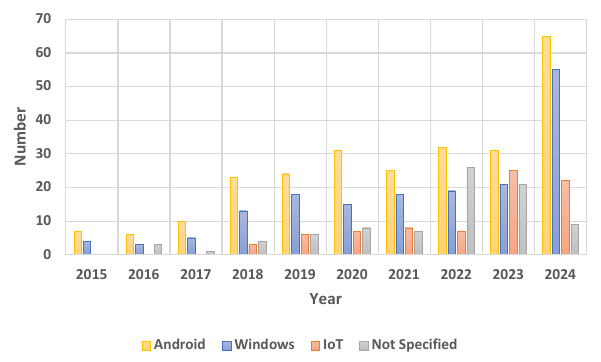}
    \caption{Research trends in target operating systems.}
    \label{Bar chart by year}
\end{figure}

The prevalence of malware targeting specific operating systems is largely influenced by the popularity and user base of these platforms. Over the years, Windows, Android, and IoT systems have become prime targets due to their widespread adoption. This has driven extensive research into malware detection and classification for these platforms. Figure \ref{Bar chart by year} illustrates the distribution of research efforts over various types of operating system based on the same survey sources used for Figure \ref{Pie Chart 2}.
\subsubsection{Windows Malware}
Windows, as the dominant operating system for personal computers, remains a prime target for malware due to its widespread use in home and enterprise environments. Its extensive adoption in critical sectors amplifies the potential impact of successful attacks, making it a persistent focus for cybercriminals \cite{Gaffney2013,Nachiyappan2020}.

\begin{figure}[h]
    \centering
    \includegraphics[width=0.6\linewidth]{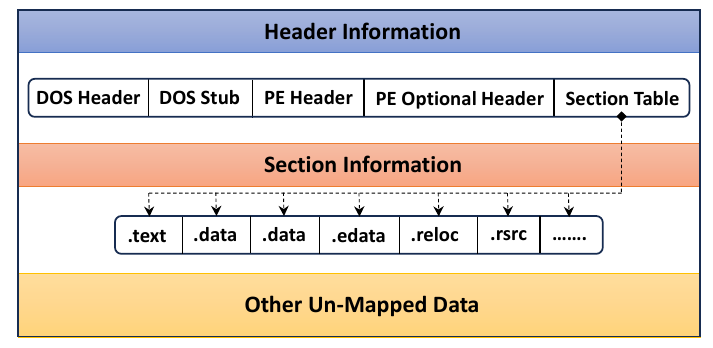}
    \caption{The structure of a PE file inspired by \cite{PEFormatMSDN2002}.}
    \label{fig:pe_structure}
\end{figure}

A central feature of Windows is the Portable Executable (PE) file format, the standard for executables, DLLs, and object code. The PE format structures critical information required by the Windows loader to manage program execution efficiently, consisting of headers, sections, and unmapped data. As illustrated in Figure \ref{fig:pe_structure}, this format ensures compatibility across Windows versions and hardware configurations, supporting executable code and associated data storage \cite{PEFormatMSDN2002}.

The header section begins with the DOS Header and DOS Stub for backward compatibility, followed by the PE Header and Optional Header, which contain vital metadata like section count and memory allocation details. Sections such as .text for executable code, .data for variables, .idata for import tables, and .rsrc for resources like icons and menus demonstrate the PE format’s adaptability. Additionally, unmapped data, including debugging information, facilitates development and troubleshooting \cite{paterson1983inside}.

While the PE format is essential for legitimate software, it is also exploited by malware. Its structured layout and widespread use make it a convenient framework for malicious software like viruses, ransomware, and trojans. The self-contained nature of PE files allows malware to embed all components necessary for attacks, evading detection and simplifying deployment. Security analyses, such as those in \cite{ling2023adversarial}, highlight how malware leverages the PE format’s capabilities to propagate and execute malicious tasks effectively.

Malware creators exploit the flexibility and robustness of the PE structure to conceal payloads, evade detection, and execute operations stealthily. This underscores the critical need for advanced detection mechanisms and user awareness to mitigate the risks posed by PE-based malware. For further insights into these challenges and countermeasures, readers can explore \cite{souppaya2013guide,zeidanloo2010all}.

\subsubsection{Android Malware}
The Android operating system has become a prime target for mobile malware due to its open-source nature, significant market share, and flexible architecture, which allow cybercriminals to exploit its vulnerabilities. As the most widely used platform for smartphones and tablets, Android powers billions of devices globally, making it an appealing platform for attackers. Its open-source codebase enables cybercriminals to analyze and exploit its features, while its flexible application management policies, including the ability to install apps from third-party sources, further expose users to risks. These factors, combined with relatively lenient app vetting processes on certain app stores, create opportunities for malicious software to infiltrate devices. Malware targeting Android devices serves diverse purposes, from monitoring user behavior and stealing sensitive data to executing unauthorized financial transactions and deploying advanced threats like ransomware, spyware, and trojans. The growing complexity of Android malware underscores the need for robust security measures, user awareness, and advanced detection techniques to protect the platform’s extensive user base.

The diversity and sophistication of Android malware continue to rise, posing significant challenges for users and security professionals alike. Common forms of malware include spyware, which monitors user activities and steals sensitive information; ransomware, which locks or encrypts data and demands payment for access; and banking trojans, which aim to extract financial credentials. An illustrative example is the UAPUSH spyware, which has infected countless devices by stealing critical user information and enabling further attacks such as fraud and identity theft \cite{Pierazzi2020}. The rapid evolution of mobile malware is evidenced by reports such as a 49 percent increase in new malware families within a single quarter in 2013 \cite{Gaffney2013}. Countering these threats requires a multifaceted approach that incorporates enhanced detection techniques, rigorous app vetting processes, user education, and proactive security measures. The collaborative efforts of researchers, platform developers, and security providers remain vital to safeguarding the Android ecosystem against this evolving threat landscape.

\subsubsection{IoT Malware}
The rapid proliferation of Internet of Things (IoT) devices has introduced significant cybersecurity challenges, particularly with the emergence of IoT-targeted malware. Over the past five years, researchers have increasingly focused on this growing threat due to the unique vulnerabilities inherent in IoT ecosystems \cite{CHENG2023267,graph2vec2,graph2vec3,GIN1,FCG_12}. IoT devices, ranging from smart home systems to industrial sensors, often lack robust security measures and operate on lightweight, resource-constrained architectures, making them prime targets for attackers. Unlike traditional systems, IoT devices are frequently deployed with default credentials, outdated firmware, or inadequate patch management, which attackers exploit to gain unauthorized access. IoT malware such as Mirai and its variants exemplify the scale of this threat, orchestrating large-scale DDoS attacks by compromising millions of devices worldwide \cite{Mirai}.

IoT malware is particularly concerning due to its ability to adapt and exploit the interconnected nature of IoT networks. Attackers leverage these devices to form botnets, enabling malicious activities such as spamming, data theft, or even ransomware attacks targeting critical infrastructures \cite{Nguyen2020IoTBotnet}. The decentralized and heterogeneous nature of IoT ecosystems complicates detection and mitigation efforts, as devices often communicate over diverse protocols. Researchers have emphasized the need for tailored detection strategies, including lightweight intrusion detection systems and ML-based anomaly detection, to address IoT-specific challenges. As IoT adoption continues to grow, securing these devices against malware threats remains a critical priority to ensure the security and safety of IoT-based applications \cite{graph2vec2,FCG_17}.

\section{Feature Engineering}

Once a sufficient collection of benign and malicious samples has been gathered and appropriately labeled, the next crucial step involves conducting feature engineering on the entire dataset prior to feeding samples into ML models. Feature engineering tries to uncover the intrinsic characteristics of files or programs that are most pertinent for distinguishing between malicious and legitimate software, subsequently generating corresponding numerical representations.

\subsection{Main Features for Malware Detection}
\begin{figure*}[h]
    \centering
    \includegraphics[width=0.8\linewidth]{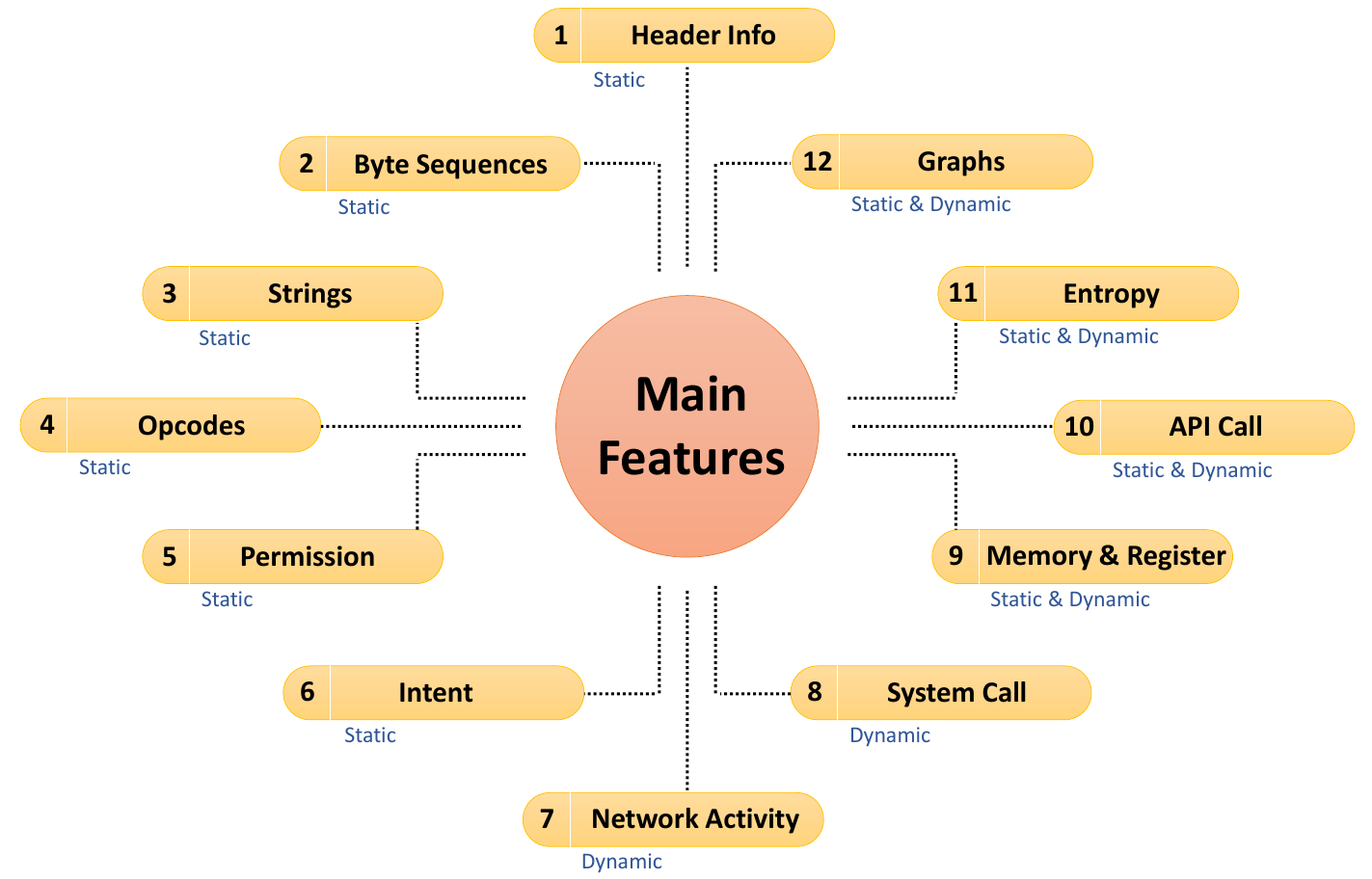}
    \caption{Main features for malware detection.}
    \label{Features}
\end{figure*}

This section provides an overview of the main features of files that can be identified through the static and the dynamic analysis methods aiming at detecting malware. Figure \ref{Features} illustrates twelve main features along with the corresponding file analysis method capable of extracting each feature.

\begin{itemize}[leftmargin=*]
    \item \textbf{Header Information:} In the PE files, the header provides essential information required by the Windows OS loader for file management. Specifically, basic statistical metrics derived from the PE header, including file size, section counts, section sizes, and counts of imported or exported functions, are frequently utilized as feature representations in PE malware detection. Recently, researchers have focused on utilizing PE header information as a primary feature source for detecting malware in PE files \cite{PE_Header_0,PE_Header_1, PE_Header_2}.
    \item \textbf{Byte Sequences:} Byte sequences offer a valuable means of capturing patterns and relationships between bytes. When analyzing byte sequences within files, researchers can pinpoint common byte patterns indicative of malware. To leverage byte sequences for detecting malware in files, scholars commonly resort to ML techniques. For instance, they might convert byte sequences into images and input them into ML algorithms like CNN. This technique is extensively documented in the literature \cite{Image_1,Image_2,Image_3,Image_4,Image_5,Image_6,Image_7,Image_8,Image_9,Image_10,Image_11,Image_12,Image_13,Image_14,Image_15,Image_16}. Additionally, they can extract byte n-grams from byte sequences and input them into various RNN algorithms to enhance detection capabilities. These approaches highlight the diverse ways, in which byte sequences can be utilized to enhance malware detection efforts \cite{Byte_Sequence_0,Byte_Sequences_1}.
    \item \textbf{Interpretable Strings:} Using interpretable strings within a file as features for malware detection can be highly effective. By extracting and analyzing these strings, malware analysts can identify suspicious or malicious patterns indicative of malicious intent. For instance, malware often uses obfuscation techniques to hide its true nature, but these strings can act as clues, helping analysts to understand the malware's functionality without executing it. Additionally, by focusing on interpretable strings, analysts can create more explainable and interpretable models for detecting malware, which is crucial for understanding the reason behind detection decisions \cite{String_0,String_1}.
    \item \textbf{Opcode:} Opcode, short for ``operation code'' is a code that specifies an operation to be performed by a computer's CPU. By analyzing the Opcodes, one can identify some patterns and characteristic behaviors of malware, such as specific sequences of instructions used in common exploit techniques or malicious operations \cite{Opcode}.
    \item\textbf{Permission:} The permissions system on Android platforms is designed to protect user privacy. Its primary purpose is to require applications to request permission to access data and system functionalities, such as making calls or using the camera. Typically, these applications need to ask for a specific set of permissions. Therefore, it is essential to carefully review these requests before granting access to ensure the desired features are appropriately managed. This system not only helps in maintaining user privacy but also serves as a critical feature in malware detection. By monitoring and analyzing the permissions requested by applications, it is possible to identify potentially malicious behavior \cite{Permission}. 
    \item \textbf{Intent:} Intent is the basic communication mechanism used to exchange the inter- and intra-app messages. An intent conveys the intention of the app to perform some actions. It specifies the label for a component, its category and actions to be performed \cite{Intent_0,Intent_1}.
    \item \textbf{Network Activity:} Malware often communicates over the network to receive commands, download additional payloads, or exfiltrate data. By analyzing the network activity, we can extract features such as the destination IP addresses, port numbers, protocols used, and the frequency and timing of communications. These features can help to identify malicious behaviors, such as connections to known malicious servers, unusual communication patterns, or attempts to exploit network vulnerabilities. Analyzing network activities can provide valuable insights into the behavior of malware and help detect previously unseen threats \cite{Network_Activity_0,Network_Activity_1}.
    \item \textbf{System Call:} These calls are requests made by a program to the operating system kernel, allowing it to perform essential tasks such as file operations, network communication, and process management. Malware often exhibits characteristic patterns of system call usage, such as attempting to access restricted resources or perform unauthorized actions. By analyzing system calls, security analysts can uncover malicious behaviors and identify potential threats \cite{System_call_0,System_Call_1}.
    \item \textbf{Memory and Registers:} Analyzing how a program interacts with memory and registers can reveal suspicious behavior indicative of malware. For instance, malware might attempt to write to sensitive memory regions, manipulate registers in unconventional ways, or use memory allocation functions in ways that deviate from normal software practices. These behavioral anomalies can be used as features in ML models aiding in the classification of files as benign or malicious based on their behavior during the static analysis \cite{Memory_Register_0,Memory_Register_1}.
    \item \textbf{API Call:} API calls represent interactions with the operating system and external libraries, offering insight into the functionality and behavior of the executable. Malware often exhibits distinct patterns of API calls, such as accessing system resources, manipulating files, or establishing network connections, which can be used to identify malicious behaviors. Analyzing API calls can reveal the intent and functionality of the malware, enabling security analysts to detect and classify threats effectively. By comparing API call patterns against known malware signatures or behavioral profiles, security tools can identify and mitigate potential threats posed by malicious files \cite{API_Call_0,ACG_1}.
    \item \textbf{Entropy:} Entropy provides a measure of randomness or uncertainty present in the file's contents. Encrypted or compressed files often exhibit higher entropy due to the increased randomness introduced by these processes. By calculating the entropy of specific regions or the entire file, analysts can detect suspicious patterns that diverse from the expected entropy levels of normal and uncompressed executables. Integrating entropy analysis into malware detection systems enhances their ability to identify potentially malicious files based on their structural characteristics \cite{Entropy_0,Entropy_1}.
    \item \textbf{Graphs:} Graph-based features play a pivotal role for malware detection by capturing the intricate relationships and structures within executable files \cite{Graph_0}. These features represent programs as graphs, where nodes can symbolize various elements such as functions, basic blocks, or system calls, and edges denote the interactions or control flow between these nodes. By analyzing the graphical representation, it becomes easier to detect patterns and anomalies that are indicative of malicious behaviors. Examples of graph-based features include control flow graphs (CFGs), function call graphs (FCGs), and system call graphs. The primary objective of this survey is to explore graph-based malware detection with a focus on explainability. Consequently, the remainder of this survey will discuss and elaborate on the most significant research in this field.
\end{itemize}
\subsection{Graph Types for Malware Detection}
Recent studies indicate that graph-based methods can effectively characterize malware behaviors and achieve high prediction accuracy. Their significance lies in the ability to offer a detailed view of program execution, aiding analysts in understanding program logic, spotting vulnerabilities, and uncovering malicious activities such as hidden or obfuscated code. Initially, this method was costly and time-consuming due to the need for meticulous analysis by skilled security experts and limited automation capabilities. However, recent advancements in ML and data analysis have revolutionized graph analysis for malware detection, enabling an automated, accurate, and cost-efficient analysis. The key graph structures utilized for malware detection are as follows:
\subsubsection{Control Flow Graph (CFG)}
It is a representation of the paths that a program can take during its execution. CFG can be extracted by reverse engineering a program's binary code. It is a directed graph, where nodes represent basic blocks of the code, and edges represent the flow of control between these blocks. Each basic block typically contains a sequence of instructions that are executed sequentially. Researchers have extensively utilized CFGs for malware detection \cite{graph2vec,GCN8,GAT1,CFG_1,CFG_2,CFG_3,CFG_4,CFG_5,CFG_6,CFG_7,CFG_9,CFG_10,CFG_12,CFG_14,CFG_15,CFG_16,CFG_17,CFG_18,CFG_19,CFG_20,CFG_21,CFG_22}. The utilization of CFGs for malware detection can be categorized into static or dynamic, depending on the methodology and nature of the features extracted from CFGs.
\subsubsection{Function Call Graph (FCG)}
It is also a type of directed graph. In this graph, nodes represent individual functions within a program, and the directed edges between nodes represent the flow of control between functions, indicating which functions call other functions. Numerous studies \cite{Multi-GNN2,GCN9,graph2vec2,Autoencoder3,GCN4,GCN5,FCG_1,FCG_2,FCG_3,FCG_4,FCG_5,FCG_8,FCG_12,FCG_13,FCG_15,FCG_16,FCG_17} utilize FCGs for malware detection and classification. By constructing an FCG, analysts can identify patterns of behavior that are characteristic of malware. For instance, malware often exhibits polymorphic or metamorphic behaviors, where the code changes its appearance to evade detection. FCGs can help in detecting such behaviors by identifying functions that have multiple call sites or that exhibit unusual control flow patterns. 
\subsubsection{API Call Graph (ACG)}
It is another form of directed graphs that can be derived from CFGs. In these graphs, the nodes signify the API calls made by the program, and the edges indicate the control flow between these calls. Malicious programs often exhibit distinct API call patterns, such as frequent access to sensitive system resources, unusual file operations, or network communication activities. By comparing the API call graph of a program to known benign and malicious patterns, one can effectively detect and classify malware \cite{graph2vec1,SAGE2,SAGE3,Multi-GNN1,GCN3,GCN7,GCN6,ACG_1,ACG_2,ACG_9,ACG_11}.
\subsubsection{System Call Graph (SCG)}
It is captured through dynamic analysis and illustrates the interactions between a program and an operating system via system calls. In this graph, nodes represent the individual system calls made by the program, and edges depict the sequence or control flow between these calls. Analyzing SCGs can help in detecting malware by identifying abnormal patterns or sequences of system calls that deviate from typical benign behaviors, thereby uncovering malicious activities \cite{SCG_1,GIN1}.
\subsubsection{Heterogeneous Graphs (HGs)} 
These are graphs that incorporate multiple types of nodes and edges to represent diverse elements and their relationships within a program. These graphs are used to capture a more comprehensive view of the program's structure and behavior by integrating different kinds of information. This capability has recently attracted significant attention from researchers, who are leveraging the advantages of heterogeneous graphs in the field of malware detection \cite{RW9,GAT2,GAT5,GAT6}.
\subsubsection{Other Graphs}
This group includes network flow graphs, program dependency graphs, and opcode graphs that are also leveraged for the sake of malware detection, though they are less commonly utilized compared to the aforementioned types of graphs. Network flow graphs represent the data flow within a network, illustrating how data packets travel between nodes \cite{nettwork_flow_graph_1,nettwork_flow_graph_2}. Program dependency graph is a graphical depiction of the source code. In this graph, vertices represent programming expressions, variables, conditions, and method calls. The edges illustrate the program and control dependencies between these vertices, showing how different parts of the code are interconnected \cite{program_dependence_graph}. Opcode graphs focus on the sequences of operation codes within the program, highlighting patterns in the executed instructions \cite{Opcode_graph_1,Opcode_graph_2,RW11}. While these graphs provide valuable insights into specific aspects of a program's behavior and can be instrumental in detecting malware, they are not as widely adopted as CFGs or FCGs for detecting malware.

\begin{figure}[h]
    \centering
    \includegraphics[width=0.5\textwidth]{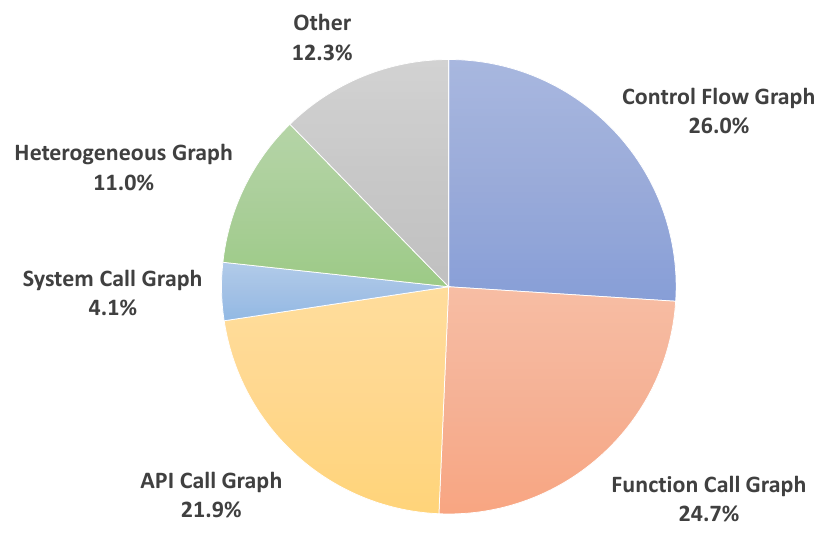}
    \caption{Distribution of graph structure used for malware detection.}
    \label{Pie Chart 1}
\end{figure}

Figure \ref{Pie Chart 1} presents the distribution of graph structures used in the literature for malware detection and classification in terms of percentage, depicted in a pie chart, utilizing the same survey sources referenced in Figure \ref{Bar chart by year}. The pie chart shows the prevalence of different graph types used for malware detection. CFGs lead at 26\%, emphasizing their importance in tracing program flows. FCGs follow closely with 24.7\%, crucial for analyzing program interactions. ACGs represent 21.9\%, showcasing their utility in monitoring API-level interactions. SCGs, at 4.1\%, and HGs, at 11\%, are used less extensively. The ``Other" category, comprising 12.3\%, reflects a diverse range of additional graph-based methods.

\section{Graph Reduction}

The initial versions of generated graphs used in malware detection, such as CFGs or FCGs, are often exceedingly large and complex. This complexity complicates the process of analyzing and understanding malware behavior and patterns due to the vast number of nodes and edges in the graphs. Furthermore, directly handling these extensive graphs during embedding and training sessions for tasks such as classification or clustering is computationally expensive and may lead to suboptimal performance. For instance, a CFG generated from a PE file contains extensive information about basic blocks, control flow edges, entry/exit points, among others. However, from a security perspective for malware detection, not all of these information are necessary. By eliminating unimportant parts, i.e., nodes or edges, the analysis can be streamlined and more efficient. Additionally, the process of simplifying CFGs must be carried out meticulously to preserve crucial information necessary for distinguishing between malware and benign CFGs. Blindly simplifying the graph may reduce its size, but it can pose challenges for efficient detection of potential malicious sub-graphs among benign CFGs. Without careful consideration during graph reduction, essential features might be lost, making it difficult for the method to distinguish between malicious and benign samples.
\begin{figure*}
    \centering
    \includegraphics[width=0.8\linewidth]{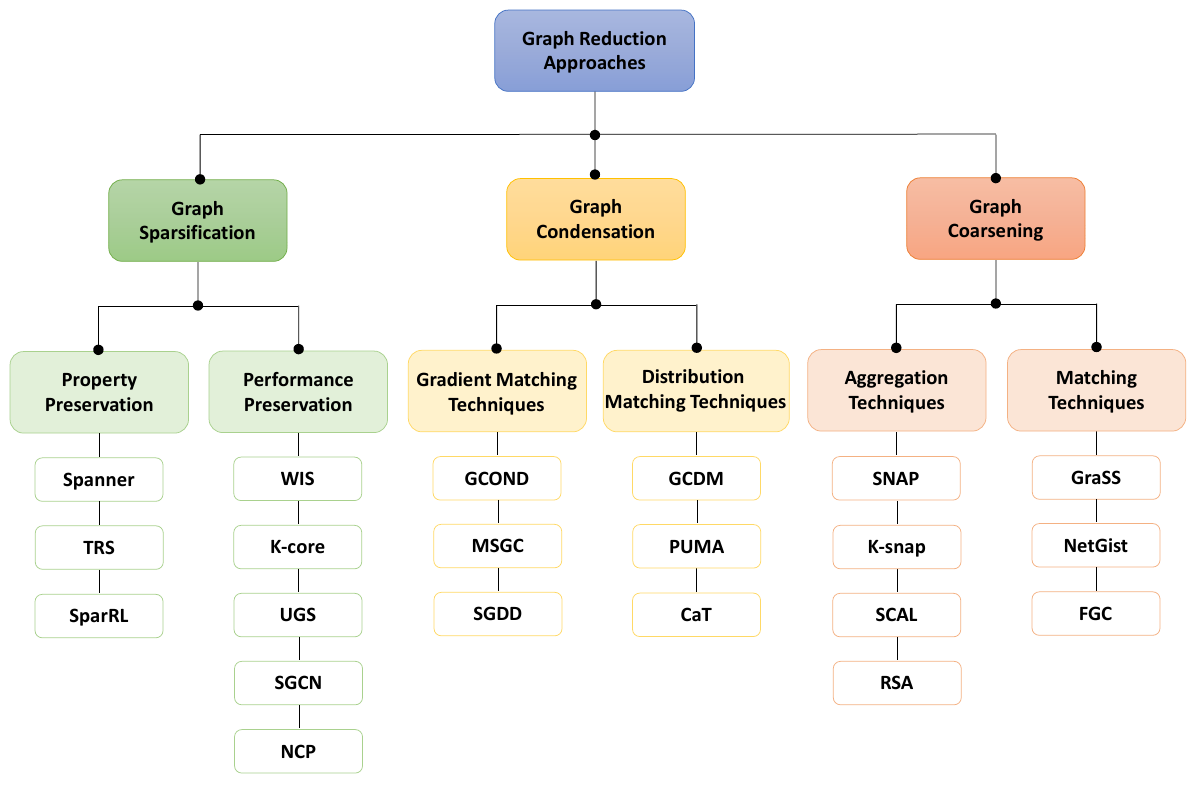}
    \caption{Taxonomy of graph reduction techniques.}
    \label{Graph_reduction_taxonomy}
\end{figure*}

In the following, we will discuss various graph reduction techniques, focusing on reducing the complexity of graphs commonly used in malware detection while ensuring that the most significant properties relevant to identifying malicious patterns are preserved. Figure \ref{Graph_reduction_taxonomy} provides a taxonomy of the main categories of graph reduction techniques, along with their main strategies and significant techniques.

\subsection{Graph Sparsification}
\begin{figure}[h]
    \centering
    \includegraphics[width=0.7\linewidth]{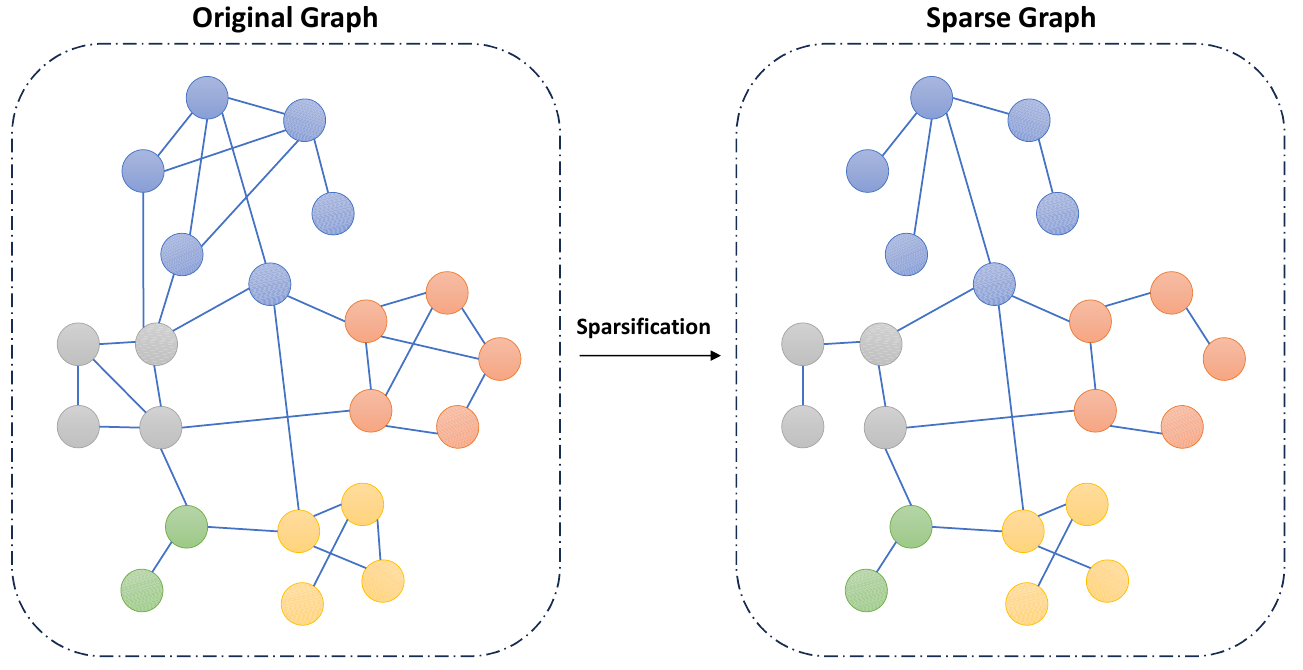}
    \caption{Graph sparsification based on the edge pruning (emphasizing on preservation of the bridge edges).}
    \label{Graph_sparsification}
\end{figure}

Graph sparsification primarily aims to create a sparse approximation by selecting essential edges (E) or nodes (V) based on specific criteria. For instance, given an original graph $G = (V, E)$, a graph sparsification algorithm aims to select existing nodes or edges from $G$ to create an sparsified graph $G' = (V', E')$, where $V'$ and $E'$ are subsets of $V$ and $E$, respectively. Conventional graph sparsification approaches have traditionally focused on preserving graph properties. However, with emerging technologies in graph learning, such as GNN models, which can effectively perform embedding and classification simultaneously, the impact of sparsification on classification performance becomes important \cite{CoreWalk}. Consequently, we study various graph sparsification methods, classifying them into those focused on preserving graph properties and those preserving performance. Figure \ref{Graph_sparsification} illustrates an example of graph sparsification based on edge pruning. In this figure, the reduction strategy emphasizes preserving bridge edges, which are edges that, if removed, would split the graph into two parts.

\subsubsection{Preserving graph properties}

Preserving graph properties aims to maintain certain structural characteristics of the original graph, while reducing its size or complexity. The term properties here refers to graph connectivity, degrees of nodes, clustering coefficients, or spectral properties. Sparsification methods iteratively sample sub-graphs to minimize a loss value measuring the approximation to the original graph. A reduced graph is called a spanner if it maintains pairwise distances, and a sparsifier if it preserves cuts or spectral properties. Althöfer et al. \cite{Spanner} proposed a spanner technique designed for weighted graphs. Beginning with an empty graph on the original node set, this method selectively adds edges from the original graph only if their weight is less than the current distance between their connected nodes in the reduced graph. 

Batson et al. \cite{TRS} proposed the Twice Ramanujan Sparsifier (TRS), which reduces the number of edges in a graph to at most half the number of nodes. This method decomposes the graph into high-conductance sub-graphs, calculates effective resistances, and samples edges based on these resistances. The sampled edges are then re-weighted. Similar to previous spanner techniques, the TRS is primarily designed for weighted graphs. Wickman et al. \cite{SparRL} proposed SparRL, a new sparsification framework based on deep Reinforcement Learning (RL). Their approach uses an RL algorithm to create a simplified version of a graph by iteratively pruning edges. The method randomly samples subgraphs, calculates node degrees, evaluates edge importance using a learned function, and prunes the least important edges.

\subsubsection{Preserving graph performance}
Preserving performance in graph sparsification aims to maintain the prediction accuracy of GNNs despite structural reductions in the graph. Techniques such as PageRank \cite{page1999pagerank}, NCP \cite{NCP}, KCenter \cite{shmoys1997approximation}, and GNN explanations \cite{ying2019gnnexplainer} are employed to select top-ranked nodes or edges. For instance, Razin et al. \cite{WIS} introduced Walk Index Sparsification (WIS), which removes edges based on Walk Index values to maintain interaction modeling. Brandeis et al. \cite{brandeis2020graph} proposed the Core-Adaptive Random Walk method (K-core), which adjusts random walks based on core indexes for link prediction tasks, enhancing training efficiency. Chen et al. \cite{chen2021unified} developed Unified GNN Sparsification (UGS), incorporating regularization during training to prune insignificant connections, reducing both edges and GNN parameters. Li et al. \cite{li2022graph} proposed the Sparsified Graph Convolutional Network, which uses a neural network graph sparsifier for edge pruning, balancing classification accuracy and graph efficiency to optimize computational performance.

Considering these two sparsification approaches for graph reduction, we believe that models focused on preserving graph performance are better suited for reducing graphs used in malware detection. These models can maintain performance in identifying malicious samples among benign ones. This preservation of performance encompasses various aspects such as maintaining key properties, including connectivity and node degrees, while effectively capturing important graph features essential for malware detection. 

\subsection{Graph Condensation}
Unlike traditional unsupervised techniques for node grouping and edge pruning, graph condensation reduces the number of nodes by learning synthetic nodes and connections in a supervised manner \cite{GCOND}. It constructs a synthetic graph \( S = \{A', X', Y'\} \) with \( N' \) nodes from the original graph \( T = \{A, X, Y\} \) with \( N \) nodes, where \( N' \ll N \). Here, \( A \in \mathbb{R}^{N \times N} \) and \( A' \in \mathbb{R}^{N' \times N'} \) are adjacency matrices, \( X \in \mathbb{R}^{N \times d} \) and \( X' \in \mathbb{R}^{N' \times d'} \) are node feature matrices, and \( Y \in \mathbb{N} \) and \( Y' \in \mathbb{N} \) are node labels in the original and synthetic graphs, respectively. The objective is to train a GNN on the smaller graph \( S \), while maintaining comparable classification performance to a model trained on the larger graph \( T \). This is formulated as a bi-level optimization problem \cite{GCOND}:
\begin{equation}
\left\{
\begin{aligned}
    & \min_{S} \quad \mathcal{L}(\text{GNN}_{\theta_S}(A, X), Y) \\
    & \text{s.t.} \quad \theta_S = \arg \min_{\theta} \mathcal{L}(\text{GNN}_{\theta}(A', X'), Y')
\end{aligned}
\right.
\end{equation}
where, \( \text{GNN}_\theta \) denotes a GNN parameterized by \( \theta \), \( \theta_S \) represents the parameters of the model trained on \( S \), and \( \mathcal{L} \) is the loss function (e.g., cross-entropy) measuring prediction discrepancies.

\begin{figure}[h]
    \centering
    \includegraphics[width=0.7\linewidth]{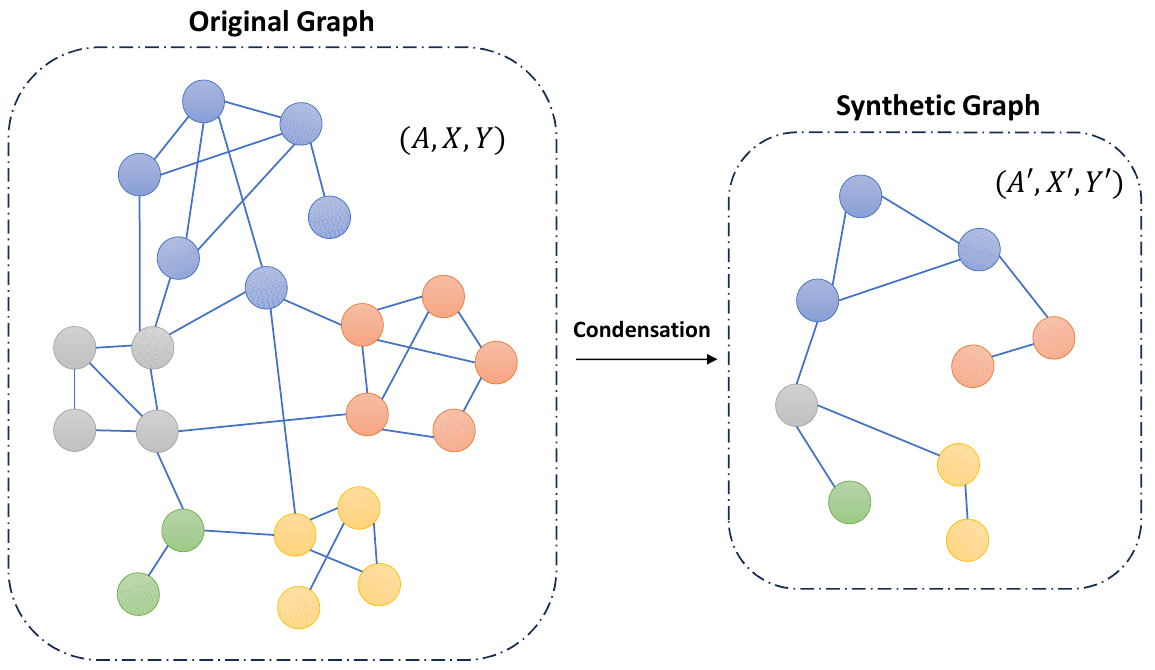}
    \caption{Graph condensation process: transforming a complex graph into its condensed form.}
    \label{Graph_Condensation}
\end{figure}

This approach distills knowledge from a large training dataset into a compact synthetic dataset, enabling comparable model performance. Figure \ref{Graph_Condensation} illustrates an example of the graph condensation approach. Graph condensation techniques primarily fall into two categories: Gradient Matching-based Techniques and Distribution Matching-based Techniques, each providing distinct methodologies to achieve the condensation objective.
\subsubsection{Gradient Matching-based Techniques} 
Graph condensation techniques that are based on gradient matching align the gradients (model updates) of GNN parameters between condensed and original graphs to replicate the training trajectory of the original graph. Jin et al. \cite{GCOND} introduced graph condensation (GCOND), the first gradient-matching-based graph condensation technique. This method optimizes a gradient matching loss to mimic the GNN training trajectory while condensing node features and structural information. GCOND demonstrated its effectiveness by generating a synthetic graph of only 154 nodes from an original graph with 153,932 nodes, achieving comparable performance across multiple GNN variants, including Graph Convolutional Network (GCN), Simplified Graph Convolutional Network (SGCN), Approximate Personalized Propagation of Neural Predictions (APPNP), and GraphSAGE.

Gao and Wu \cite{MSGC} proposed Multiple Sparse Graphs Condensation (MSGC), which condenses the original graph into multiple small and sparse graphs, while evenly distributing connections among nodes of different classes. This ensures that the sparsified graphs reflect the original class distribution. Yang et al. \cite{SGDD} introduced Structure-broadcasting Graph Dataset Distillation (SGDD), which optimizes node features through gradient matching and addresses Laplacian Energy Distribution (LED) shift using an LED Matching strategy based on an optimal transport distance. Mao et al. \cite{mao2023gcare} proposed Graph Condensation through Adversarial Regularization (GCARe) to ensure fairness in distilling node subgroups. In GCARe, condensing GNNs act as a generators in a Generative Adversarial Network (GAN), producing node embeddings, while a discriminator predicts subgroup memberships, ensuring fair subgroup representation during graph condensation.
\subsubsection{Distribution Matching-based Techniques}
Unlike gradient matching-based approaches, distribution matching-based techniques align the statistical distributions of probabilities across all nodes or intervals in the graph, without relying on gradient information. Liu et al. \cite{GCDM} proposed Graph Condensation through receptive field Distribution Matching (GCDM), a technique that preserves structural and feature information by iteratively aligning the embeddings of the original and condensed graphs. GCDM alternates between updating node features and graph generator parameters, followed by minimizing embedding discrepancies.

Liu et al. \cite{PUMA} introduced the PsUdo-label guided Memory bAnk (PUMA) for continual graph learning, which condenses incoming graphs by refining node features and labels using MLP and GNN encoders. PUMA minimizes embedding discrepancies with a Maximum Mean Discrepancy (MMD) loss, accommodating both labeled and unlabeled nodes while balancing historical and new graphs. Similarly, Liu et al. \cite{CaT} proposed the Condense and Train (CaT) model, which uses a Condensed Graph Memory (CGM) to align synthetic and original data distributions via MMD. CaT addresses imbalanced learning by storing condensed graphs instead of entire incoming graphs.

A notable limitation of most graph condensation techniques is their reliance on node labels for supervision during synthetic graph construction. As such, these methods are primarily evaluated on labeled datasets like Cora, Citeseer, Arxiv, Genius, Flickr, and Reddit. However, this dependence limits their application in scenarios like malware analysis, where graphs (e.g., CFGs) are labeled as benign or malware, but individual nodes lack specific labels, making such techniques less effective.

\subsection{Graph Coarsening}
\begin{figure}[h]
    \centering
    \includegraphics[width=0.7\linewidth]{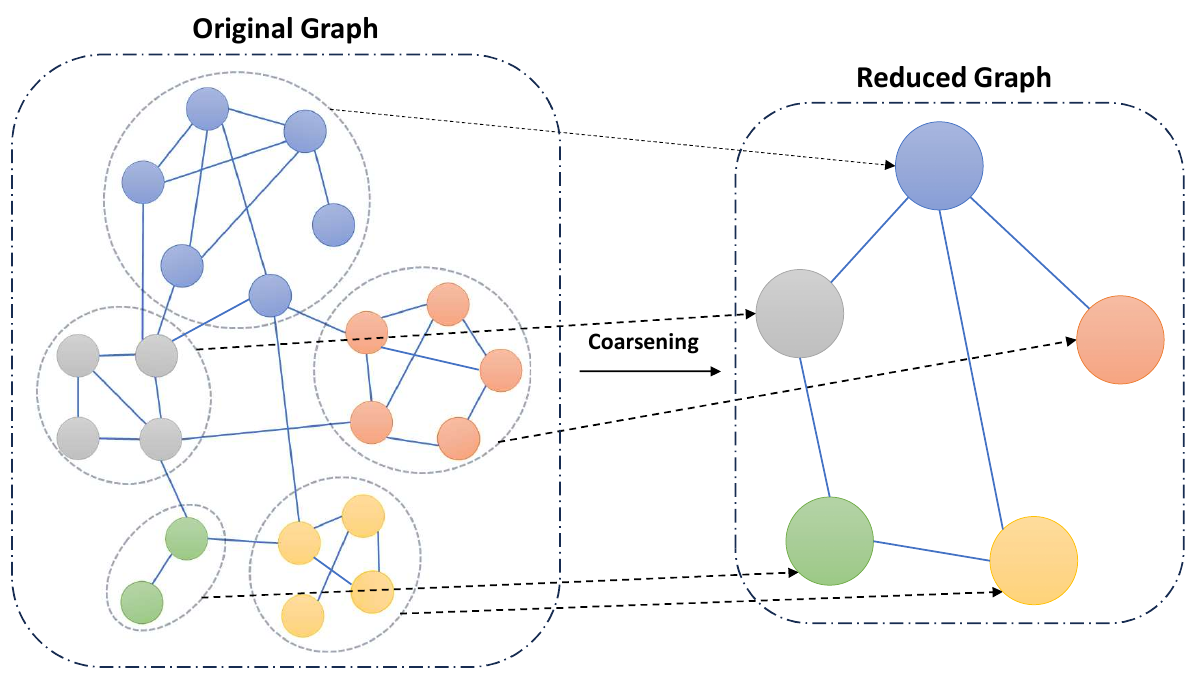}
    \caption{Graph coarsening process: transferring a complex graph into its reduced form.}
    \label{Graph_Coarsening}
\end{figure}

Graph coarsening is an unsupervised graph reduction approach that groups nodes into super-nodes and aggregates their edges into super-edges, resulting in a coarsened graph. Unlike graph condensation, which requires supervision and node labels for training synthetic graphs, graph coarsening can be applied to graphs without node labels. This makes it particularly suitable for scenarios involving unlabeled graphs, such as CFGs and FCGs used in malware analysis. Figure \ref{Graph_Coarsening} illustrates an example of the coarsening process, where nodes on the left are grouped into super-nodes on the right. The two main categories of graph coarsening are aggregation-based techniques and matching-based techniques.

\subsubsection{Aggregation-Based Techniques}
Graph coarsening using aggregation-based approaches reduces graphs in a reconstruction-free manner by creating super-nodes and super-edges based on node attributes and their relationships.

Tian et al. \cite{NetGist} proposed two techniques, SNAP and K-snap, for graph summarization. SNAP groups nodes based on user-selected attributes and their relationships, iteratively refining these groups until they are fully connected or disconnected. K-snap extends SNAP to create a summary graph with a specified number of groups (\(k\)), splitting groups iteratively to preserve the graph's structure while managing noise and missing data.
Huang et al. \cite{SCAL} introduced Scaling up GNNs via graph coarsening (SCAL) for semi-supervised node classification in attributed graphs. SCAL computes super-node feature vectors as the mean of constituent nodes' features and assigns labels based on the majority class. GNNs trained on the coarsened graph use the optimized parameters for predictions, enhancing scalability.
Loukas et al. \cite{RSA} proposed Restricted Spectral Approximation (RSA), which guarantees spectral and cut accuracy by using Restricted Spectral Similarity (RSS). RSA approximates the Laplacian matrix of the original graph through multi-level coarsening, progressively reducing the Laplacian matrix until a target size or an error threshold is reached. Additionally, they proposed single-level coarsening by local variation, selecting subsets, and adjusting the Laplacian size based on predefined criteria.

\subsubsection{Matching-Based Techniques}
Graph coarsening using matching-based approaches reduces graphs by iteratively merging nodes based on matching criteria such as similarity or optimization objectives, creating super-nodes, while preserving essential structures and relationships.

\begin{figure*}[h]
    \centering
    \includegraphics[width=\linewidth]{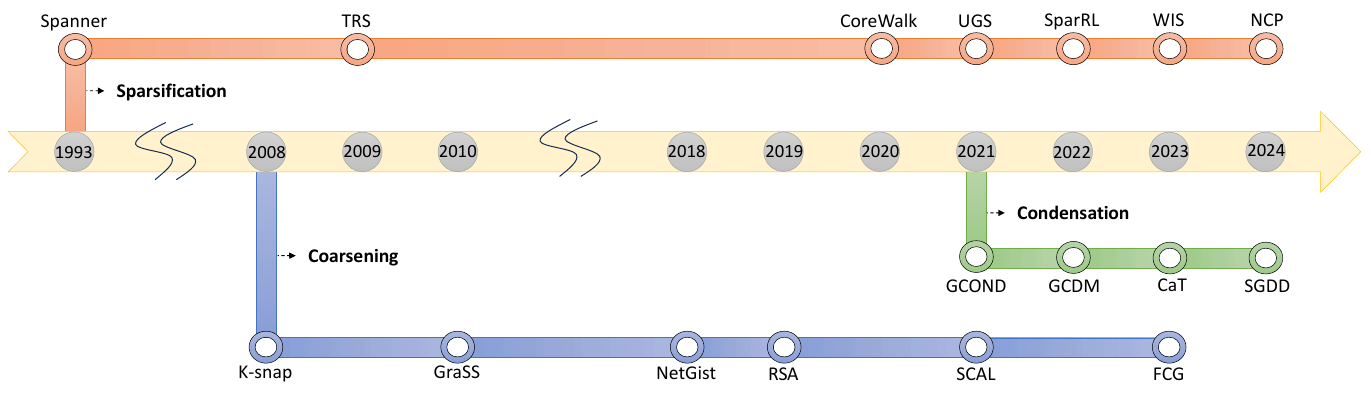}
    \caption{Timeline of graph reduction techniques, representing most significant works of each category.}
    \label{Graph_reduction_timeline}
\end{figure*}

LeFevre and Terzi \cite{lefevre2010grass} proposed Graph Structure Summarization (GraSS), which summarizes graph structures for query answering using a random-worlds framework. GraSS employs a greedy algorithm that iteratively merges pairs of nodes to minimize an objective function, stopping when \(k\) super-nodes are formed or no further improvements can be made. The method addresses graph structure queries such as adjacency, degree, and eigenvector centrality. Amiri et al. \cite{amiri2018netgist} introduced NetGist, which uses deep Q-learning to merge nodes until the graph reaches a desired summary size. The process iteratively reduces the graph over multiple episodes by merging nodes, ensuring the reduced graph contains fewer nodes than a specified fraction of the original graph. A divide and conquer strategy evaluates summary quality, while Q-learning parameters are refined to improve subsequent results. Kumar et al. \cite{FGC} proposed Featured Graph Coarsening (FGC), a technique with similarity guarantees. FGC iteratively updates variables to optimize an objective function through gradient calculations and matrix operations. The process continues until a stopping criterion is met, producing a coarsened Laplacian matrix optimized for graph-based problems with user-defined parameters.
Figure \ref{Graph_reduction_timeline} illustrates the timeline of key advancements in graph sparsification, condensation, and coarsening techniques, highlighting their respective years of origin. 

In summary, this section explored three graph reduction approaches—graph sparsification, condensation, and coarsening—along with their subcategories, each suited to specific domains and objectives. In malware analysis, a key challenge is the absence of individual node labels, as CFGs typically assign a single binary label (malware or benign) to the entire graph. Since nodes often share uniform characteristics derived from program flows, structure, and execution paths, unsupervised graph reduction strategies are recommended. Techniques such as graph sparsification with performance preservation, reconstruction-free graph coarsening, and clustering-based methods effectively address this uniformity while preserving critical properties for malware detection.

\subsection{Graph Reduction Strategies for Malware Detection}
In this section, we outline the criteria used to recommend graph reduction methods suitable for malware detection. Our approach was driven by a focus on methods that are both explainable and capable of performing graph classification. Specifically, we recommended only those methods that meet the following criteria:

\begin{enumerate}
    \item The method must be applicable for graph classification. Given that malware detection often requires a comprehensive understanding of the entire graph structure, methods that can classify graphs as a whole are more suitable for this task.
    \item The method must not rely on node labeling in its reduction technique. Methods that do not require node labels during reduction are generally more flexible and can be applied in scenarios, where node labels are unavailable or unreliable. This characteristic enhances the method's applicability to various datasets, including those used in malware detection.
    \item The method must be explainable. Explainability is crucial in cybersecurity applications, where understanding the rationale behind a classification decision is as important as the decision itself. Explainable methods allow practitioners to trace back the steps leading to a decision, which is critical in validating the results and gaining insights into the nature of the detected malware.
\end{enumerate}

Based on these criteria, the following five methods were identified as recommended for malware detection:

\begin{itemize}[leftmargin=*]
    \item \textbf{TRS}: This method is recommended because it effectively sparsifies the graph, while preserving essential spectral properties, making it suitable for graph classification. TRS does not depend on node labeling and is inherently explainable due to its reliance on well-understood spectral techniques.
    \item \textbf{WIS}: It excels in link prediction and graph classification tasks, leveraging walk indices to preserve key interaction information within the graph. It is explainable through the interpretation of walk-based metrics and does not require node labeling, making it a robust choice for malware detection.
    \item \textbf{SGCN}: SGCNs are included due to their effectiveness in node and graph classification. SGCN reduces computational complexity, while maintaining classification accuracy, and its explainability stems from the use of convolutional operations, though it typically uses node labels. However, its strong performance in graph classification justifies its inclusion.
    \item \textbf{SCAL}: SCAL focuses on scalable graph clustering, while preserving the structure necessary for classification. Although it utilizes node features, it provides a balance between performance and explainability, particularly in scenarios, where graph partitioning and summarization are critical.
    \item \textbf{FGC}: This method is particularly well-suited for influence-based coarsening and graph partitioning. FGC’s explainability is derived from its focus on retaining influence relationships, making it a valuable tool in identifying significant components within the graph that correlate with malware behaviors.
\end{itemize}
Table \ref{tab:my-table} provides a comprehensive comparison of various graph reduction methods, highlighting their suitability for different tasks, including malware detection, based on key criteria such as performance preservation, explainability, and supervised or unsupervised operations.
\begin{table*}[h]
\centering
\caption{Main graph reduction techniques and their characteristics (NC: Node Classification, LP: Link Prediction, GC: Graph Classification, CD: Community Detection, GS: Graph Summarization, SC: Spectral Clustering, QA: Query Answering, R: Recommend, and NR: Not Recommend).}
\label{tab:my-table}
\resizebox{\textwidth}{!}{%
\begin{tabular}{@{}lcc cccccc@{}}
\toprule
\multirow{2.5}{*}{\textbf{Methods}} & \multicolumn{2}{c}{\textbf{Pruning}} & \multirow{1.5}{*}{\textbf{Performance}} & \multirow{1.5}{*}{\textbf{Property}} & \multirow{1.5}{*}{\textbf{Supervised/}} & \multirow{2.5}{*}{\textbf{Task}} &\multirow{2.5}{*}{\textbf{Explainable}} & \multirow{2.5}{*}{\textbf{Malware Detection}} \\
\cmidrule(r){2-3}
 & \textbf{Nodes} & \textbf{Edges} & \textbf{Preserving}&\textbf{Preserving} &\textbf{Unsupervised} & & & \\ 
\midrule
Spanner\cite{Spanner}  & No    & Yes   & No    & Yes   & U   & CD  & Yes  & NR  \\
TRS \cite{TRS}     & No    & Yes   & No    & Yes   & U   & GC  & Yes  & R   \\
SparRL \cite{SparRL}  & No    & Yes   & No    & Yes   & U   & LP  & Yes  & NR  \\
WIS  \cite{WIS}    & No    & Yes   & Yes   & No    & U   & LP,GC & Yes  & R   \\
NCP  \cite{NCP}    & No    & Yes   & Yes   & No    & U   & GC  & Yes  & R   \\
CoreWalk \cite{CoreWalk} & No    & Yes   & Yes   & No    & U   & LP  & Yes  & NR  \\
UGS \cite{UGS}     & No    & Yes   & Yes   & No    & U   & NC,GC & Yes  & R   \\
SGCN \cite{SGCN}    & No    & Yes   & Yes   & Yes   & S   & NC  & No   & NR  \\
GCOND \cite{GCOND}   & Yes   & Yes   & Yes   & No    & S   & NC  & No   & NR  \\
MSGC  \cite{MSGC}   & Yes   & Yes   & Yes   & No    & S   & NC  & No   & NR  \\
SGDD   \cite{SGDD}  & Yes   & Yes   & Yes   & No    & S   & NC  & No   & NR  \\
GCDM  \cite{GCDM}   & Yes   & Yes   & Yes   & No    & S   & NC  & No   & NR  \\
PUMA  \cite{PUMA}   & Yes   & Yes   & Yes   & No    & S   & NC  & No   & NR  \\
CaT   \cite{CaT}   & Yes   & Yes   & Yes   & No    & S   & NC  & No   & NR  \\
SNAP  \cite{snap}   & Yes   & Yes   & Yes   & Yes   & U   & GS  & Yes  & NR  \\
K-snap \cite{snap}  & Yes   & Yes   & Yes   & Yes   & U   & GS  & Yes  & NR  \\
SCAL   \cite{SCAL}  & Yes   & Yes   & Yes   & Yes   & S   & NC,GC & Yes  & R   \\
RSA    \cite{RSA}  & Yes   & Yes   & Yes   & Yes   & U   & SC  & Yes  & NR  \\
GraSS \cite{lefevre2010grass}   & Yes   & Yes   & Yes   & Yes   & U   & QA  & Yes  & NR  \\
NetGist \cite{NetGist} & Yes   & Yes   & Yes   & Yes   & S   & CD  & Yes  & NR  \\
FGC    \cite{FGC}  & Yes   & Yes   & Yes   & Yes   & U   & NC,GC & Yes  & R   \\ 
\bottomrule
\end{tabular}
}
\end{table*}

It is important to note that our recommendations are strongly influenced by the need for explainability in the methods. In the context of malware detection, being able to explain why a method arrived at a particular decision is as important as the accuracy of the decision itself. By prioritizing methods that offer clear, understandable results, we ensure that the outcomes of the detection process can be trusted and verified, which is essential in cybersecurity applications.

\section{Graph Embedding}
\begin{figure}[h]
    \centering
    \includegraphics[width=0.8\linewidth]{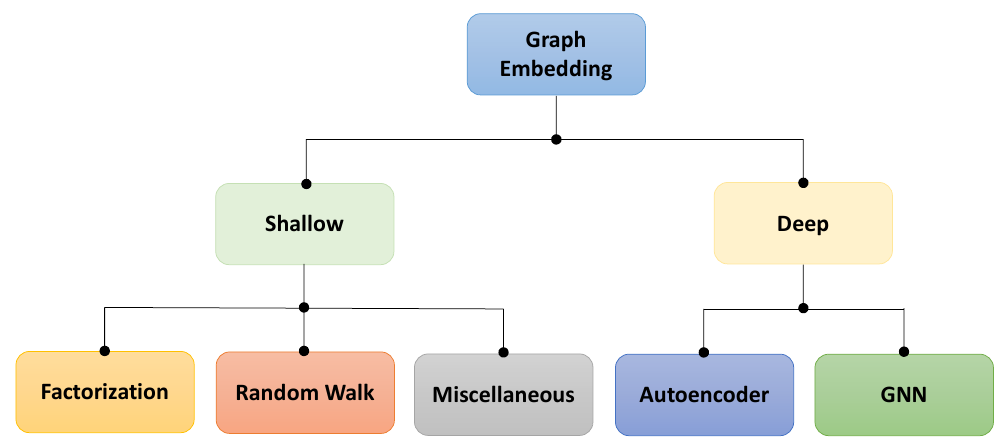}
    \caption{Taxonomy of graph embedding techniques.}
    \label{Graph embedding}
\end{figure}

Graph embedding is a technique in the analysis of graph-structured data, aiming to transform high-dimensional and complex graph data into low-dimensional vector representations, while preserving the intrinsic properties and relationships of the original graph. This transformation facilitates the application of various ML algorithms that require fixed-size input, enabling tasks such as node classification, link prediction, and graph classification. Essentially, graph embedding techniques can be divided into two main categories: shallow and deep embeddings. Figure \ref{Graph embedding} illustrates a taxonomy of the graph embedding techniques from the literature, distinguishing between shallow and deep embedding methods. This figure highlights the most significant contributions within these two groups.

\subsection{Shallow Embedding Techniques}
These techniques seek to represent graph data in a reduced-dimensional vector space, ensuring that nodes, which are adjacent in the graph remain proximate in this compressed space, thus maintaining the structural relationships among nodes. Below, we provide a detailed analysis of these approaches, categorizing them into methods based on matrix factorization, methods deriving from random walks, and miscellaneous methods.
\begin{itemize}[leftmargin=*]
    \item \textbf{Matrix factorization-based approach:} This traditional approach relies on matrix factorization and typically follows a two-step process. First, a matrix based on the node proximity is created, where each matrix element quantifies the closeness between two nodes within the graph. In the second step, a dimensionality reduction method is applied to this matrix to produce the embedding for the nodes. Figure \ref{factorization} illustrates the general idea of the matrix factorization approach for graph embedding, where $[x_{c1}, ..., x_{cm}]$ represents the low-dimensional embedding of the target node. Numerous methodologies have been proposed in the scholarly literature to enhance matrix factorization-based graph embedding techniques, such as Locally Linear Embedding (LLE) \cite{LLE}, Laplacian Eigenmaps (LE) \cite{LE}, Graph Factorization (GF) \cite{GF}, GraRep \cite{GraRep}, and High-Order Proximity preserved Embedding (HOPE) \cite{HOPE}. Among these, one of the more recent advancements is HOPE, which specifically targets the unique challenges associated with the embedding of directed graphs by concentrating on the preservation of asymmetric transitivity, thus addressing a crucial aspect of directed relationships within graph structures. This property, critical in directed graphs, highlights the likelihood of a direct edge from one node to another if a directed path exists between them, capturing the directed edges' asymmetric relationships. HOPE achieves this by approximating high-order proximities, such as Katz and Rooted PageRank, which inherently consider the directionality of edges. It utilizes a dual embedding strategy, where each node is represented by both a source and a target vector. The embedding vectors are learned through a scalable matrix factorization approach, specifically a generalized singular value decomposition (SVD) that approximates high-order proximity measures without the computationally intensive task of direct proximity matrix computation.
    \begin{figure*}[h]
    \centering
    \includegraphics[width=\textwidth]{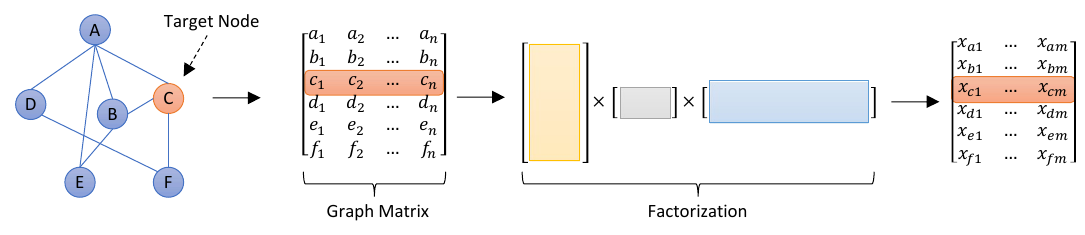}
    \caption{Factorization-based embedding.}
    \label{factorization}
    \end{figure*}
    \item \textbf{Random walk-based approach:} The main concept of this method involves generating random walks between nodes in a graph to encapsulate its structural features (see Figure \ref{random walk}). As a result, nodes that frequently appear together in short random walks are likely to have similar embeddings. After generating these random walks, the skip-gram model is employed to enhance the node embedding process. This model predicts surrounding node contexts within each walk, effectively capturing the structural relationships within the graph. Unlike the static proximity measures used in traditional matrix factorization-based methods, this technique leverages co-occurrence during random walks as an indicator of node similarity. This more adaptable approach has shown encouraging results in a range of applications. DeepWalk \cite{DeepWalk} pioneered the use of random walks on graphs to generate extensive sequences of nodes, which are treated as sentences and processed using Word2Vec to enhance the probability of accurately predicting the node context given a target node. This approach, compared to matrix factorization techniques, offers significantly lower time complexity, making it well-suited for large-scale graph representation learning. However, DeepWalk primarily focuses on local node relationships, which complicates the process of determining the most effective random walk sampling sequences. Another significant method based on random walks is Node2Vec \cite{Node2vec}, which serves as an advanced version of DeepWalk. Node2vec employs the parameters $p$ and $q$ to bias the behavior of its random walks. The parameter $p$ influences the random walk's propensity to revisit nodes $U$ that it has previously visited, with lower values of $p$ increasing this likelihood. The parameter $q$, on the other hand, supports inward and outward explorations: a $q$ value greater than 1 encourages the walk to explore nodes nearer to $U$, whereas a $q$ less than 1 favors distant nodes. This mechanism essentially seeks to strike a balance between Depth-First Search (DFS) and Breadth-First Search (BFS), enhancing the algorithm's ability to capture the graph's topological nuances.
    Starting from node $U$ and currently at node $W$, the algorithm gives node $S_2$ a weight of $1$, noting that its distance from node $U$ is the same as that from node $W$ to node $U$. For node $S_1$, which is closer to node $U$, a weight of $1/p$ is assigned, while node $S_3$, being further away, receives a weight of $1/q$. Subsequently, these weights are normalized, and probabilities for the next step of the walk are assigned to each node based on these normalized weights.
    \begin{figure*}[h]
    \centering
    \includegraphics[width=0.65\textwidth]{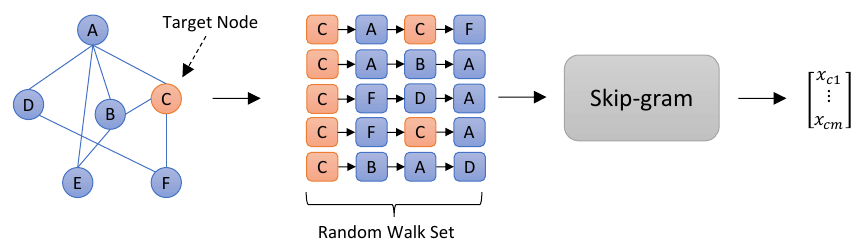}
    \caption{Random walk-based embedding.}
    \label{random walk}
    \end{figure*}
Additionally, methods such as HARP \cite{Harp}, Struct2Vec \cite{struct2vec} and Metapath2Vec \cite{metapath2Vec} have been developed to expand on random walk-based techniques.

Random walk-based techniques have been applied to various fields, including malware detection \cite{RW1,RW2,RW3,RW4,RW6,RW7,RW8,RW9,RW11}. A notable recent contribution in this area is API2Vec, as detailed in \cite{RW8}. API2Vec introduces a graph-based API embedding method to enhance malware detection by addressing the interleaving of API calls in multi-process malware. This method constructs a dual graph model comprising a Temporal Process Graph (TPG) and Temporal API Graphs (TAGs) within each process. These graphs capture the complex inter-process and intra-process relationships and behaviors. Through heuristic random walks over these graphs, API2Vec generates paths representing detailed behavior patterns of malware. These paths are then vectorized using the Doc2Vec algorithm to create numerical embeddings for both individual API calls and their execution contexts, significantly improving the detection accuracy and robustness of malware detection systems.
\item \textbf{Miscellaneous methods:} Large-scale Information Network Embeddings (LINE) \cite{LINE} does not utilize random walks, but is often compared with random walk-based methods. Figure \ref{LINE} Shows the process of graph embedding utilizing LINE. This technique preserves both local and global network structures by employing first-order and second-order proximities, making it suitable for capturing the direct and indirect relationships between nodes. First-order similarity refers to the likeness between two directly linked nodes, illustrated by their joint probability distribution. In contrast, second-order similarity focuses on the count of shared neighboring nodes, assessing how similar the local neighborhood structures (contexts) of the nodes are. LINE addresses the challenge of high-dimensional data by reducing dimensionality, which facilitates easier visualization and analysis of complex networks. Graph2Vec \cite{graph2vec} is another shallow embedding technique distinct from both factorization-based and random walk-based embedding methods. Graph2Vec utilizes the rooted subgraphs around nodes to capture the graph's topology. These rooted subgraphs are analogous to words in a document, and the embeddings are learned based on the presence and arrangement of these subgraphs, reflecting the structural properties of the graph. To address the challenge of malware detection and classification, the authors in \cite{graph2vec1,graph2vec2,graph2vec3} employed the Graph2Vec method for embedding the structural properties of the corresponding graphs.
\begin{figure*}[h]
    \centering
    \includegraphics[width=0.8\textwidth]{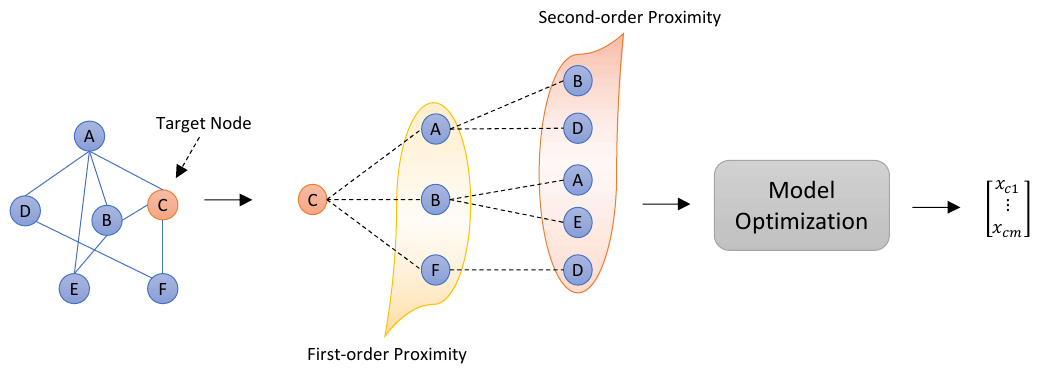}
    \caption{Graph embedding by LINE.}
    \label{LINE}
    \end{figure*}
\end{itemize}
Shallow graph embedding techniques have played a significant role in graph representation learning, but they exhibit several limitations, which necessitate more advanced deep approaches. One major drawback is that these methods often focus on local neighborhood structures without capturing the global topology of the graph. This can result in embeddings that lack important global information. Additionally, shallow embeddings do not generalize well to unseen nodes or subgraphs because they learn fixed embeddings rather than generalizable functions. These methods also face scalability issues with large graphs, due to the computational cost of processing extensive random walks or full graph structures. Lastly, shallow embedding techniques typically generate static embeddings and do not adapt to changes in the graph over time without complete re-training. These limitations highlight why more deep embedding methods have become more prevalent in graph-based malware detection tasks.
\subsection{Deep Embedding Techniques}
These techniques represent a significant advancement over shallow embedding methods by offering more powerful representations of the graph structures. These techniques enable the extraction of high-level features and facilitate the embedding of more abstract representations of nodes and their interactions. As a result, deep graph embeddings can achieve superior performance on tasks like node classification and graph clustering, where capturing sophisticated structural details is crucial. The deep embedding techniques can be broadly categorized into two main types: autoencoder-based and GNN-based methods. Each category utilizes different architectures and mechanisms to learn effective graph representations.
\begin{itemize}[leftmargin=*]
    \item \textbf{Autoencoder-based methods:} The main concept behind this approach is to use a deep neural network architecture that consists of two primary components: an encoder and a decoder. Figure \ref{autoencoder-based} depicts the main idea of autoencoder-based graph embedding. The encoder part compresses the input graph data, typically the adjacency matrix and node features, into a lower-dimensional latent space. This latent representation captures the essential information of the graph, preserving both structural and feature-based similarities. The decoder then attempts to reconstruct the original graph data from this embedded representation. The quality of the embedding is often measured by the accuracy of the reconstruction, under the assumption that a good embedding will retain all the necessary information to accurately rebuild the original graph. This method not only helps in reducing the dimensionality of the data but also in learning significant graph patterns that are useful for various downstream tasks such as malware detection, graph classification, and node clustering.
    Several studies, such as structural deep network integration (SDNE) \cite{SDNE}, DNNs for graph representation learning (DNGR) \cite{DNGR}, deep recursive network embedding (DRNE) \cite{DRNE}, and adversarially regularized graph autoencoder (ARGA) \cite{ARGA}, have adopted the autoencoder concept to address challenges for graph embedding. In the context of graph-based malware detection, \cite{RW1,RW2,Autoencoder3} utilize the autoencoder design to embed graphs. \cite{RW1} begins by extracting FCGs from analyzed programs. These graphs are then embedded into a low-dimensional feature space using the node2vec graph embedding technique. This procedure translates the structural characteristics of a FCG into a vector that effectively represents the graph’s structure, while maintaining neighborhood relations. This embedded vector is subsequently processed by a stacked denoising autoencoder (SDA), which is designed to learn a latent representation of the vector, capturing the essential features of the graph for further analysis.  
    \begin{figure*}[h]
    \centering
    \includegraphics[width=\textwidth]{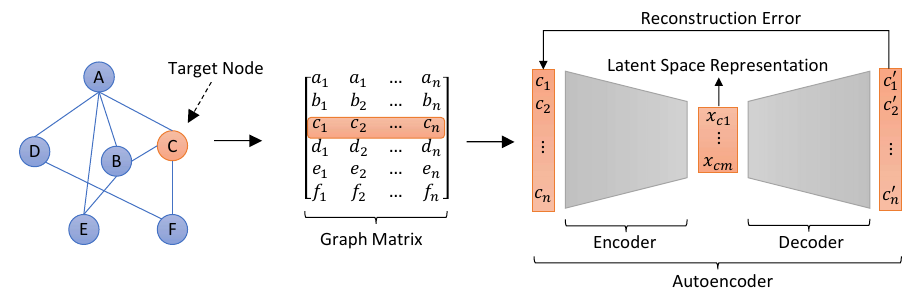}
    \caption{Autoencoder-based embedding.}
    \label{autoencoder-based}
    \end{figure*}
    \item \textbf{GNN-based methods:} They play an essential role in contemporary deep embedding techniques and are recognized as a comprehensive framework for deploying Deep Neural Networks (DNNs) on graph-structured data. The core concept behind GNNs is that the representation vectors of the nodes are influenced by both the graph's structural configuration and any features associated with these nodes. This dual dependency allows GNNs to capture complex patterns in the data, making them highly effective for performing tasks involving relational information like malware detection. The basic idea of GNN-based embedding is shown in Figure \ref{GNN-based}. Several variants and algorithms have been developed based on the foundational principles of GNNs such as GCNs \cite{GCN1,GCN2}, GraphSAGE \cite{GraphSAGE}, Graph Isomorphism Networks (GINs) \cite{GIN}, Spatial-Temporal Graph Neural Networks (STGNNs) \cite{GSTN1,GSTN2}, Graph Attention Networks (GATs) \cite{GAT} and Dynamic Graph Neural Networks (DGNNs) \cite{DGNN1,DGNN2}.
    \begin{figure*}[h]
    \centering
    \includegraphics[width=0.8\textwidth]{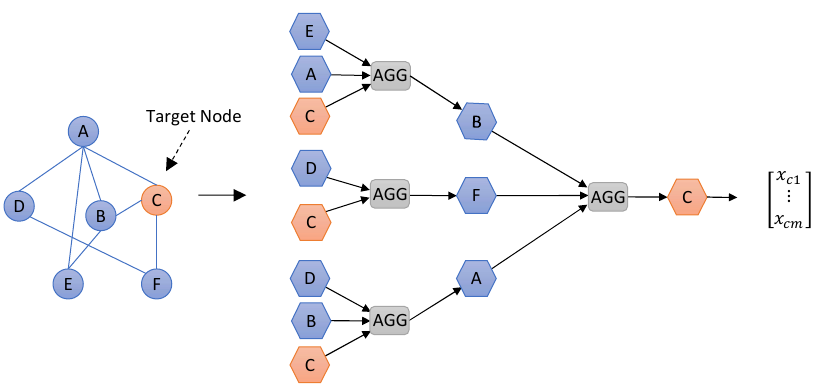}
    \caption{GNN-based embedding.}
    \label{GNN-based}
    \end{figure*}
    GCN generalizes the concept of convolution from traditional grid-like structures (such as images) to graphs, allowing them to handle non-Euclidean data. The authors in the references \cite{GCN01, GCN02, GCN3, GCN4, GCN5, GCN6, GCN7, GCN8, GCN9, GCN10, GCN11} utilized this technique to embed various types of graphs, such as CFGs and FCGs, specifically for malware detection. These graph embedding methods led to impressive performance improvements in detecting malicious software. By effectively capturing the structural information and intricate relationships within the graphs, these studies demonstrated the robustness and efficacy of graph-based approaches in enhancing the accuracy and reliability of malware detection. Additionally, the GraphSAGE method has been used to manage and analyze larger malware graphs with greater efficiency \cite{SAGE1, SAGE2, SAGE3}. Moreover, GAT introduces an attention mechanism to the GNN framework, enabling the model to assign different importance weights to different nodes in a neighborhood during the aggregation process. This attention mechanism enhances the model's ability to focus on the most relevant parts of the graph, potentially improving performance in tasks, where the importance of node connections varies. Recently, this method has garnered significant interest from researchers in the field of malware detection due to its remarkable capability and performance \cite{GAT1,GAT2,GAT3,GAT4,GAT5,GAT6}. In addition, GIN is specifically designed to effectively distinguish between graph structures and capture complex patterns. \cite{GIN1, CFG_2} demonstrated that this graph embedding method can deliver outstanding performance for malware detection and classification. The combination of GNN-based graph embeddings has been employed in some studies, \cite{Multi-GNN1, Multi-GNN2}, in order to harness the unique strengths of each method and achieve superior results. In particular, \cite{Multi-GNN2} introduces Family-Aware Graph neural network (FAGnet), which integrates GCN, GraphSAGE, GAT, and GIN. The use of multiple GNN-based methods in FAGnet aims to validate its robustness and effectiveness across diverse graph neural network architectures. While these foundational GNN models have demonstrated significant success in malware detection, recent advancements have introduced more sophisticated graph-based embedding techniques designed to address specific challenges such as feature propagation, robustness, and dynamic node representation. These include Graph Convolutional Network via Initial Residual and Identity (GCNII) \cite{GCNII}, which addresses over-smoothing with residual connections, H\textsubscript{2}GCN \cite{H2GCN}, which improves feature propagation, and Generalized PageRank Graph Neural Network (GPR-GNN) \cite{GPR-GNN}, which extends PageRank concepts for dynamic node representation. Other notable models include Equivariant Graph Neural Network (EGNN) \cite{EGNN} for equivariant learning, CPGNN \cite{CPGNN} for context preservation, and Graph Substructure Network (GSN) \cite{GSN}, which focuses on subgraph-level pattern recognition. Additionally, Gated Bi-Kernel Graph Neural Network (GBK-GNN) \cite{GBK-GNN} leverages a dual-kernel mechanism, GNN based on Aggregation Perturbation (GAP) \cite{GAP} enhances robustness through aggregation perturbation, and Spatio-Spectral Graph Neural Network (S\textsuperscript{2}GNN) \cite{S2GNN} integrates spatial and spectral processing.
\end{itemize}

\begin{figure*}
    \centering
    \includegraphics[width=\textwidth]{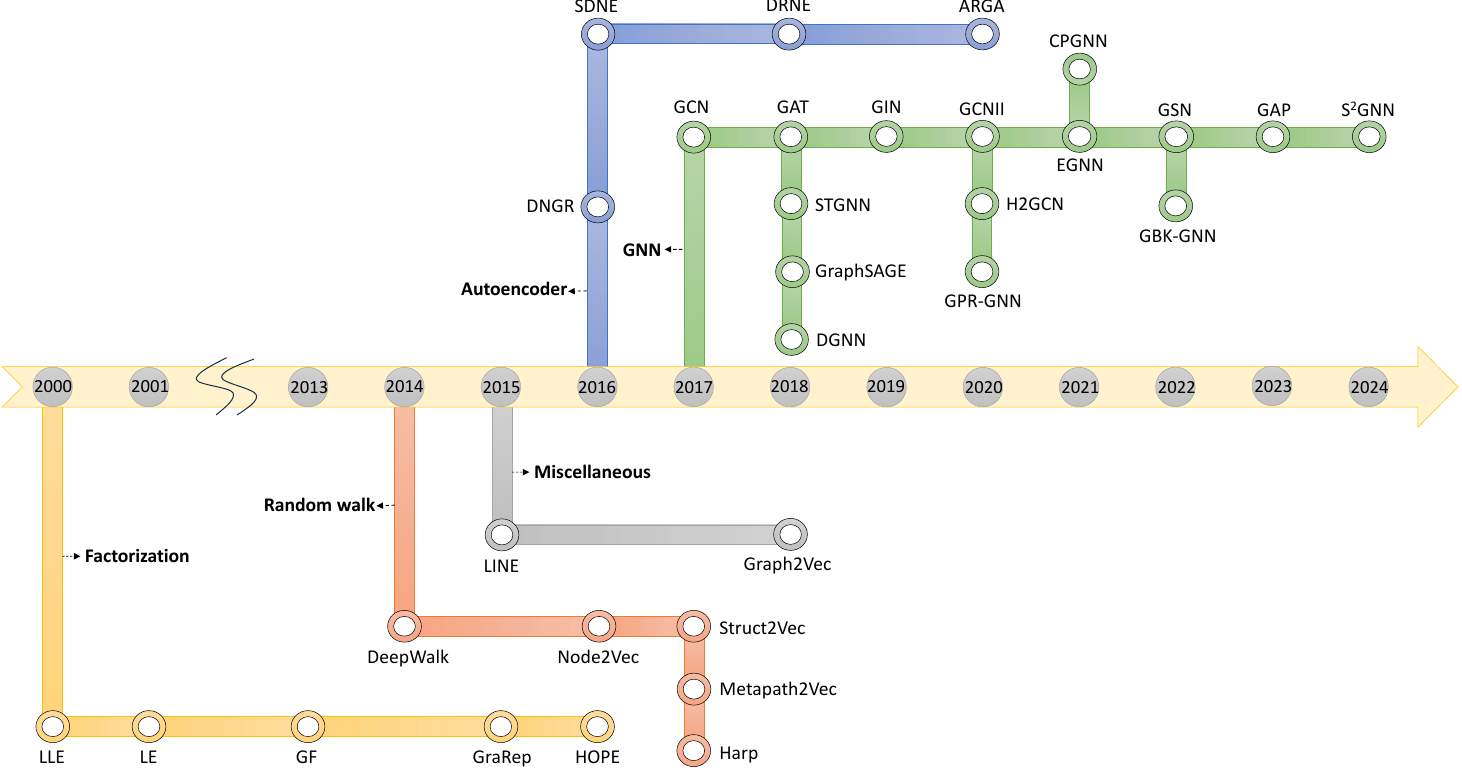}
    \caption{Timeline of graph embedding techniques, representing most significant techniques of each category.}
    \label{Embedding Timeline}
\end{figure*}

Building upon the categorization presented in Figure \ref{Graph embedding}, Figure \ref{Embedding Timeline} provides a detailed timeline of the development of graph embedding methods, showcasing the chronological progression of both shallow and deep embedding techniques. Moreover, Table \ref{embedding_comparison} provides a comprehensive comparison of graph embedding methods, highlighting their characteristics and suitability for malware detection tasks. These methods are categorized based on their underlying approaches, such as factorization, random walks, autoencoders, and GNNs. Key attributes include explainability, which is critical for understanding model decisions in security applications; robustness to obfuscation, a common technique to evade detection; computational cost and scalability, which determine the feasibility of deploying these methods in real-world. Additionally, the applicability to malware detection column rates each method's effectiveness in addressing the unique challenges of malware analysis.

\begin{table}[h]
\caption{Comparison of graph embedding methods for malware detection (R: Recommended and NR: Not Recommended).}
\label{embedding_comparison}
\centering
\begin{tabular}{p{2.5cm} p{1.7cm}<{\centering} p{1.8cm}<{\centering} p{1.5cm}<{\centering} p{1.2cm}<{\centering} p{1.3cm}<{\centering} p{1.3cm}<{\centering}} 
\toprule
\textbf{Methods} & \textbf{Approaches} & \textbf{Explainability} & \textbf{Robustness to Obfuscation} & \textbf{Cost} & \textbf{Scalability} & \textbf{Malware Detection} \\ 
\midrule
LLE \cite{LLE} & Factorization & Medium & Low & Low & Medium & NR \\ 
LE \cite{LE} & Factorization & Medium & Low & Low & Medium & NR \\ 
GF \cite{GF} & Factorization & Medium & Low & Low & Medium & NR \\ 
GraRep \cite{GraRep} & Factorization & Medium & Low & Medium & Medium & NR \\ 
HOPE \cite{HOPE} & Factorization & Medium & Low & Medium & Medium & NR \\ 
DeepWalk \cite{DeepWalk} & Random Walk & Low & Medium & Low & High & R \\ 
Node2Vec \cite{Node2vec} & Random Walk & Low & Medium & Medium & High & R \\ 
Struct2Vec \cite{struct2vec} & Random Walk & Low & High & High & Medium & R \\ 
Harp \cite{Harp} & Random Walk & Low & Medium & Medium & High & R \\ 
Metapath2Vec \cite{metapath2Vec} & Random Walk & Low & Medium & High & Medium & R \\ 
LINE \cite{LINE} & Miscellaneous & Low & Low & Low & High & NR \\ 
Graph2Vec \cite{graph2vec} & Miscellaneous & Low & Low & Medium & Medium & NR \\ 
SDNE \cite{SDNE} & Autoencoder & Medium & High & High & Medium & R \\ 
DNGR \cite{DNGR} & Autoencoder & Medium & High & High & Medium & R \\ 
DRNE \cite{DRNE} & Autoencoder & Medium & High & High & Medium & R \\ 
ARGA \cite{ARGA} & Autoencoder & Medium & High & High & Medium & R \\ 
GCN \cite{GCN1,GCN2} & GNN & High & High & High & Low & R \\ 
GraphSAGE \cite{GraphSAGE} & GNN & High & High & High & Medium & R \\ 
GIN \cite{GIN} & GNN & High & High & High & Low & R \\ 
STGNN \cite{GSTN1,GSTN2} & GNN & High & High & High & Low & R \\ 
GAT \cite{GAT} & GNN & High & High & High & Low & R \\ 
DGNN \cite{DGNN1,DGNN2} & GNN & High & High & High & Low & R \\
GCNII \cite{GCNII} & GNN & High & High & High & Low & R \\
H\textsubscript{2}GCN \cite{H2GCN} & GNN & High & High & High & Low & R \\
GPR-GNN \cite{GPR-GNN} & GNN & High & High & High & Low & R \\
EGNN \cite{EGNN} & GNN & High & High & High & Low & R \\
CPGNN \cite{CPGNN} & GNN & High & High & High & Low & R \\
GSN \cite{GSN} & GNN & High & High & High & Low & R \\
GBK-GNN \cite{GBK-GNN} & GNN & High & High & High & Low & R \\
GAP \cite{GAP} & GNN & High & High & High & Low & R \\
S\textsuperscript{2}GNN \cite{S2GNN} & GNN & High & High & High & Low & R \\
\bottomrule
\end{tabular}%
\end{table}

\subsection{Decision Making}
The decision-making process in malware detection systems transforms the features extracted from graph embeddings into actionable classifications. This step involves utilizing a decision-making layer, typically appended to GNNs or other embedding models, to produce final predictions.

In traditional machine learning workflows, classification relies on algorithms like SVM, kNN, Random Forests, and Logistic Regression applied to precomputed graph embeddings \cite{graph2vec1,graph2vec2}. These methods operate independently from the embedding generation process and are effective for smaller, well-labeled datasets. However, their disconnection from the embedding step can limit their performance when applied to large, complex, or dynamic graphs.

In GNN-based approaches, the decision-making process is seamlessly integrated into the model \cite{GCN1}. After processing the input graph through layers of GNN, the final embeddings (typically vectors representing nodes, edges, or entire graphs) are passed to a classification layer. This layer commonly includes the following components:
\begin{itemize}[leftmargin=*]

\item \textbf{Fully Connected Networks (FCNs):} Fully connected layers serve as dense classifiers, where the graph-level or node-level embeddings are passed through one or more layers of neurons. These layers learn to map the extracted features to specific classes, such as benign or malicious. FCNs provide flexibility in learning non-linear decision boundaries, making them a preferred choice for complex malware detection tasks.

\item \textbf{Softmax Activation:} For multi-class classification tasks, a softmax layer is used at the output to normalize the scores into probabilities for each class. This enables intuitive interpretations of the classification output, with the highest-probability class being chosen as the prediction. For example, a malware detection model may assign probabilities to classes like benign, ransomware, or trojan, guiding analysts toward actionable decisions.

\item \textbf{Binary Classifiers:} In binary malware detection scenarios, where the goal is to distinguish between benign and malicious samples, a sigmoid activation function is often used instead of softmax. This simplifies the decision-making layer while ensuring accurate probability-based outputs for threshold-based decisions.

\end{itemize}

\section{Explainability}
ML models have shown significant promise across various domains, including medicine, healthcare, cybersecurity, spam and malware detection, and critical infrastructures~\cite{duddu2018survey,Mahdavifar}. Their integration into high-stakes applications has necessitated a deeper understanding of their decision-making processes and the influence of AI on outcomes~\cite{goodman2017european}. This has led to a growing demand for transparency and explainability in AI, especially in critical contexts like medicine, where detailed justifications are crucial for informed decisions~\cite{preece2018stakeholders, tjoa2020survey}. Key terms associated with understanding AI decisions include Understandability, comprehensibility, interpretability, explainability, and transparency~\cite{arrieta2020explainable}, with understandability being closely tied to transparency.

The goals of XAI depend on its audience and their specific needs. For instance, developers and product owners require explainability to identify model limitations and improve functionality, while end-users, such as medical professionals or cybersecurity experts, need it to trust the model's predictions and make informed decisions. In malware detection, explainability supports tasks like understanding detection mechanisms, evaluating performance, investigating incidents, and adapting defense strategies.

The main goals of XAI can be categorized as follows:

\begin{itemize}[leftmargin=*]
    \item \textbf{Trustworthiness:} Confidence in the model’s predictions and behavior when deployed~\cite{ribeiro2016should}.
    \item \textbf{Causality:} Identifying cause-and-effect relationships between data variables~\cite{cheng2021causal}.
    \item \textbf{Transferability:} Improving or adapting the model for new problems~\cite{szegedy2013intriguing}.
    \item \textbf{Informativeness:} Providing insights into how decisions are made internally.
    \item \textbf{Conﬁdence:} Ensuring the model's reliability and stability~\cite{yu2013stability}.
    \item \textbf{Fairness:} Avoiding bias or favoritism in decision-making~\cite{mehrabi2021survey}.
    \item \textbf{Accessibility:} Allowing non-expert users to engage with the model's processes.
    \item \textbf{Interactivity:} Enabling user interaction and feedback.
    \item \textbf{Privacy Awareness:} Preventing privacy breaches from internal model representations.
\end{itemize}

While goals like fairness and privacy awareness are more relevant to human-centered or privacy-sensitive applications, the primary explainability goals for malware detection are trustworthiness, causality, informativeness, and confidence. These aspects ensure that malware detection models can be reliably used in security-critical environments and that users can trust and confidently deploy them.

\begin{figure}[h]
    \centering
    \includegraphics[width=0.8\linewidth]{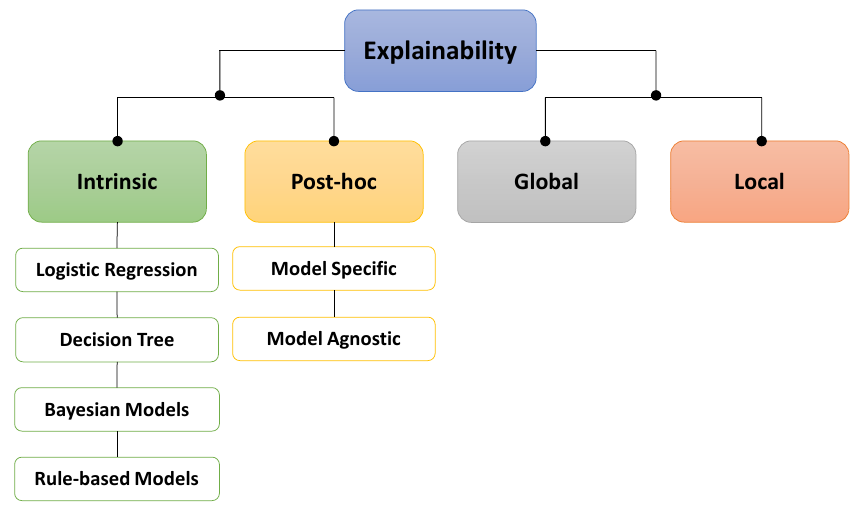}
    \caption{General taxonomy of XAI approaches.}
    \label{fig:exp_cat}
\end{figure}

After identifying the purpose and beneficiaries of explainability, it is essential to explore how explainability can be achieved. As illustrated in Figure~\ref{fig:exp_cat}, the explainability of AI models can be categorized from different perspectives. One common categorization distinguishes between local and global explainability. 
Local explainability focuses on understanding a specific decision or prediction with respect to its input. It aims to reveal the causal relationships between the input and the corresponding output, thereby fostering trust in individual predictions. In contrast, global explainability examines the model as a whole, including its parameters, structures, and learned representations. This approach provides insights into the overall mechanism and functioning of the model, helping users build trust in the model itself.

Using another approach, the AI models can be divided into two broad categories concerning explainability: Intrinsic (i.e., transparent models) and Post-hoc explainability. Some models are intrinsically explainable and provide the user with information on how they make their decisions with different levels of transparency. On the other hand, some models are opaque, and the relation between the input and the output is invisible to the users. These types of models need some post-processing or interpretability techniques to provide some explanations about their decisions to the end users.

\subsection{Intrinsic Explainability}
Transparent models inherently offer a degree of explainability and interpretability. According to the literature, these models can be categorized into three hierarchical levels: simulatability, decomposability, and algorithmic transparency, where each level encompasses the characteristics of the preceding one.

Simulatability refers to the ability of a model to be understood and mentally simulated by humans. Model complexity plays a key role here, as higher complexity makes it harder to grasp the model purely through cognitive efforts. 
Decomposability requires that all components of a model—including inputs, calculations, and parameters—are understandable. A decomposable model allows users to interpret its processes without external tools or explanations.
Algorithmic transparency ensures that the process by which a model produces an output for a given input is clear and understandable to the user, qualifying it as transparent. Examples of such models include linear/logistic regression, decision trees, k-NN, rule-based models, and Bayesian models.

\subsection{Post-hoc Explainability}
\begin{figure}[h]
    \centering
    \includegraphics[width=0.4\linewidth]{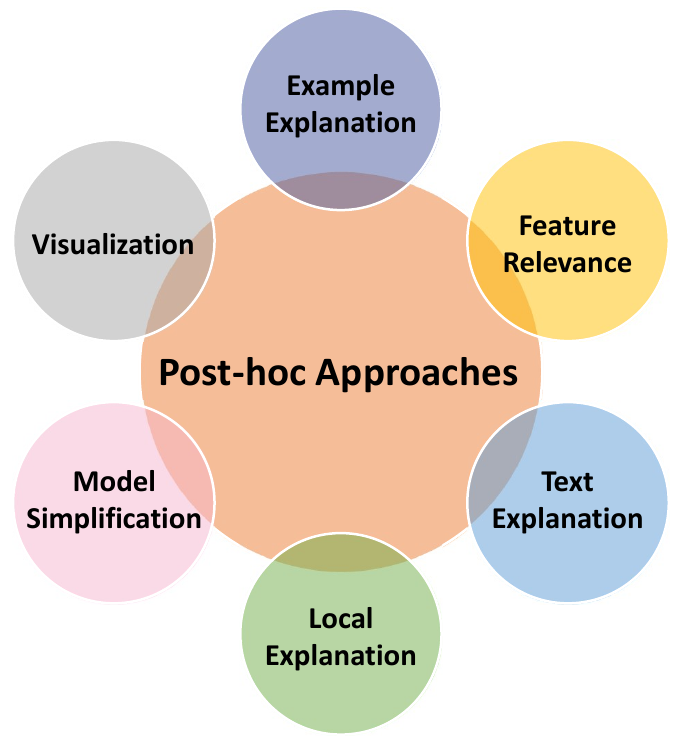}
    \caption{Post-hoc explainability approaches.}
    \label{fig:exp_app}
\end{figure}

Post-hoc explainability focuses on adding an additional layer of processing to explain models. Various techniques have been proposed in the literature, as illustrated in Figure~\ref{fig:exp_app}, that are inspired by how humans explain complex systems. Below are some widely used approaches:

\begin{itemize}[leftmargin=*]
    \item \textbf{Visualization:} Techniques that leverage dimensionality reduction methods to visualize a model's behavior in an interpretable form.
    \item \textbf{Text Explanation:} Generating textual descriptions to clarify model decisions.
    \item \textbf{Feature Relevance:} Calculating scores for input features to assess their impact on the model's output, enabling feature importance ranking.
    \item \textbf{Example Explanation:} Providing representative data examples to illustrate the model's decision-making process.
    \item \textbf{Local Explanation:} Breaking down the solution space into smaller regions to explain specific parts of the model’s functionality.
    \item \textbf{Model Simplification:} Constructing simpler surrogate models that approximate the behavior of complex models while maintaining similar performance.
\end{itemize}

Post-hoc techniques are essential for models that lack inherent explainability. These techniques can be divided into model-agnostic and model-specific categories, as shown in Figure~\ref{fig:exp_cat}. Model-agnostic approaches explain their predictions without delving into its inner workings. Conversely, model-specific techniques are tailored to the algorithms underlying the target model. The following subsections provide an overview of these methods.

\subsubsection{Model-agnostic}
Model-agnostic techniques can be applied to any target model and typically fall into visualization, feature relevance, and model simplification categories. 
One prominent model-agnostic technique is Local Interpretable Model-agnostic Explanations (LIME)~\cite{ribeiro2016should}, which explains individual predictions by training a linear surrogate model on data points near the desired input. Variants like Anchor~\cite{ribeiro2018anchors}, OptiLIME~\cite{visani2020optilime}, LORE~\cite{guidotti2018local}, and Anchor Local Interpretable Model-Agnostic Explanations (aLIME)~\cite{ribeiro2016nothing} have extended or improved LIME. Another rule-based simplification technique is proposed by Bastani et al.~\cite{bastani2017interpretability}, where they approximate the target model using axis-aligned decision trees.
Shapley Additive explanations (SHAP)~\cite{lundberg2017unified} is a popular feature relevance technique based on game theory, assigning importance values to features. Other game-theory-inspired approaches include~\cite{strumbelj2010efficient, chen2021explaining}. Visualization techniques based on sensitivity analysis have also been explored~\cite{cortez2011opening, CORTEZ20131}.

Model-agnostic techniques focus on simplifying, visualizing, and identifying feature relevance, providing insights without relying on the model's internal structure.

\subsubsection{Model-specific}
Model-specific explainability techniques are tailored to the internal structure and characteristics of a specific ML model, enabling more precise interpretations of its decision-making process. These techniques often leverage the unique properties of the target model to provide deeper insights. Depending on the model type, these approaches vary significantly, as lightweight models and deep learning models exhibit different levels of complexity and interpretability.

Lightweight models, such as k-NNs and decision trees, are inherently transparent and do not typically require extensive post-hoc explanations. However, models like SVMs and various deep learning architectures—Multi-Layer Perceptrons (MLPs), CNNs, RNNs, and GNNs—pose greater challenges for explainability due to their complexity. The following sections explore specific techniques for explaining these models.

\begin{itemize}[leftmargin=*]
    \item \textbf{SVM}, while not inherently transparent, can be explained using model simplification techniques. For instance, SQRex-SVM~\cite{barakat2007rule} extracts rules from support vectors using a modified sequential covering algorithm. Fu et al.~\cite{fu2004extracting} generated hyper-rectangular rules based on hyperplane intersections, and other works~\cite{nunez2002support, nunez2006rule} incorporated training data into the rule extraction process. 
    Visualization techniques have also been applied to explain SVMs, especially in domains like medicine and chemistry. For example, kernel matrix information was used to visualize SVMs behavior in~\cite{ustun2007visualisation}, and a heat map coloring technique was proposed for drug discovery tasks~\cite{rosenbaum2011interpreting}.
    \item \textbf{MLPs}, as foundational deep learning models, are often treated as black boxes due to the opacity of their hidden layers. Early rule-based approaches, such as DeepRED~\cite{zilke2016deepred}, extended the Continuous/discrete Rule Extractor via Decision tree induction (CRED) method~\cite{sato2001rule} to handle MLPs with multiple hidden layers. Among feature relevance techniques, DeepLIFT~\cite{shrikumar2017learning} decomposes predictions by propagating contributions from neurons, while Zhang et al.~\cite{zhang2018opening} used Garson's algorithm to measure the predictor importance for clinical applications.
    \item \textbf{CNNs}, due to their layered architecture, exhibit high complexity, making them difficult to explain. However, their visual data processing capabilities lend themselves to intuitive visualization techniques. Grad-CAM~\cite{selvaraju2017grad}, SmoothGrad~\cite{smilkov2017smoothgrad}, and Integrated Gradients~\cite{sundararajan2017axiomatic} highlight important input regions for predictions. Network Dissection~\cite{zhou2018interpreting} assigns semantic labels to CNN units, while rule-based methods like Discretized Interpretable Multi Layer Perceptron (DIMLP)~\cite{bologna2019simple} and decision tree-based techniques~\cite{zhang2019interpreting} simplify model behavior for interpretability.
    \item \textbf{RNN} excel at sequential data tasks but are difficult to interpret due to their ability to model long-term dependencies. Techniques like Layer-wise Relevance Propagation (LRP)~\cite{arras2019explaining} and network attractors~\cite{ceni2020interpreting} provide insights into RNNs behavior. Van Luong et al.~\cite{van2021designing} proposed an interpretable RNN architecture using deep unfolding optimization, enhancing sparse approximation and interpretability.
\end{itemize}

\subsection{GNN Explainability}
\begin{figure}[h]
    \centering
    \includegraphics[width=0.6\linewidth]{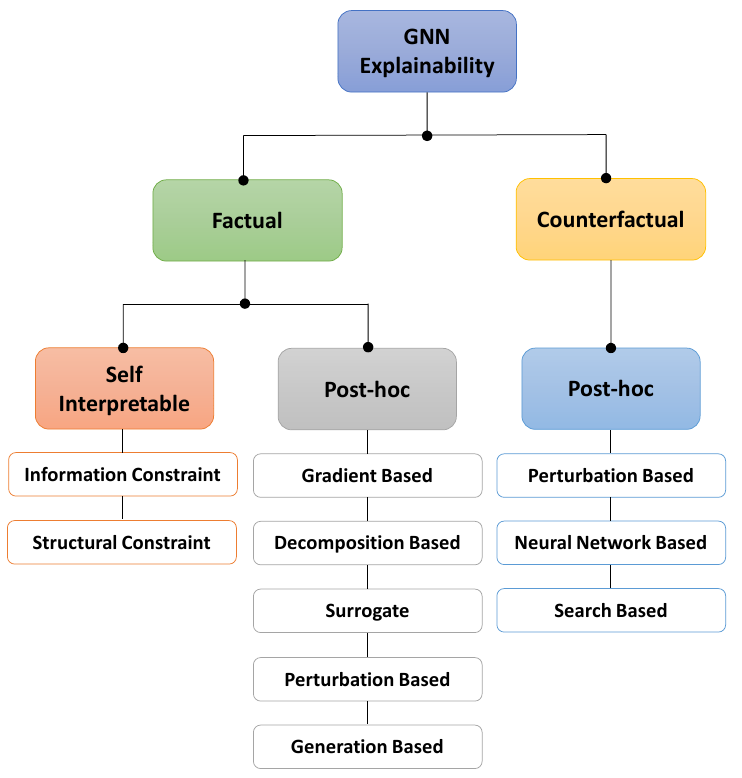}
    \caption{Taxonomy of explainable GNN approaches.}
    \label{fig:gnn_exp}
\end{figure}

Same as other types of deep learning models, GNN has received a lot of attention in recent years and has helped to solve several complex problems in different domains, including healthcare~\cite{zitnik2018modeling}, recommended systems~\cite{chen2022grease}, and fraud detection~\cite{rao2020xfraud}. Due to their high usage in critical areas, the ability to explain their decision-making process is essential. In general, the existing methods in the literature can be categorized into two broad categories of factual and counterfactual~\cite{kakkad2023survey}. Factual explanations focus on describing the features or parts of the input graph that directly contribute to the model's decision. This is about identifying which existing elements (nodes, edges, node features) in the graph are most influential for the prediction. Counterfactual explanations, on the other hand, are about speculating ``what if'' scenarios. They describe how altering the input graph could change the prediction in a specific way. The proposed taxonomy, shown in Figure~\ref{fig:gnn_exp}, utilizes the above categorization at its highest level. GNN explainers can be categorized into self-interpretable models and post-hoc. In post-hoc explainability, the process of extracting the explanation is done after the target model is trained and works as a post-processing step, while in the self-interpretable models, the explainability module is part of the model itself, and the explanation generated at the same time with the prediction. GNN explainers can also be categorized into instance-level and model-level. Instance-level explainers only provide an explanation for a single sample, while model-level explanations aim to explain the function of the whole model. Since we may have both instance-level and model-level methods in each category as presented in Figure~\ref{fig:gnn_exp}, this classification is not included in the taxonomy.

\subsubsection{Counterfactual (CF)}
These methods aim to provide explainability by detecting the smallest changes in the input that can affect the target model outputs. They can be further categorized into perturbation-based, neural network-based, and search-based.

\textbf{a) Perturbation-based:} These methods aim to create counterfactual explanations for both graph and node classification tasks by modifying the edges through either removal or addition. One of the interesting works utilizing this approach is CF-GNNExplainer~\cite{lucic2022cf} 
, where a binary matrix $P$ is used to perturb the target graph adjacency matrix and iteratively modified until the optimal counterfactual example is found. The optimization problem is defined based on two loss functions: $L_{pred}$ and $L_{dist}$, where the former is prediction loss for the counterfactual example and the latter is the distance loss between the original and the counterfactual graphs. Another significant work is Counterfactual and Factual (CF\textsuperscript{2})~\cite{tan2022learning}, which tries to combine factual and counterfactual reasonings. The factual explanation aims to find a sub-graph that is enough for making a decision, while the counterfactual method finds the necessary sub-graph. CF\textsuperscript{2} goal is to combine both of these intuitions and find a sufficient and necessary sub-graph for the target model to make the correct decision. In~\cite{chen2022grease}, authors used the same technique to propose a GNN-based Recommendation System Explainer (GREASE). The main difference between recommended systems and a regular GNN is that they rank the nodes by assigning scores instead of classifying them. Therefore, they made two modifications to the original techniques. First, they designed a loss function based on the score before and after adding the perturbation, and second, they did not perturb the whole graph.

\textbf{b) Neural Network-based:} In this group of techniques, unlike the perturbation-based techniques, a neural network is used to generate the counterfactual explanation rather than minimally perturb the input until reaching the counterfactual explanation. Robust Counterfactual explainer (RCExplainer)~\cite{bajaj2021robust}
is one of the works that follow this approach to generate robust counterfactual explanations for GNNs. RCExplainer starts with modeling the GNN's logic using several decision regions distinguished based on a set of linear decision boundaries of the target model. Then, based on these decision boundaries, a loss function is designed to be used in training a neural network that generates a small subset of edges of the input sample as a counterfactual explanation. In~\cite{ma2022clear}, authors used the graph variational autoencoders to propose a technique for generating counterfactual explanations that overcome challenges such as generalization, optimization, and causality. Using an autoencoder, they encode the input graph into a latent representation, and then a decoder generates the counterfactual explanation as a fully connected version of the input graph with different probabilities for each edge. Unlike RCExplainer, counterfactual explanation generator (CLEAR)~\cite{ma2022clear} can generate an explanation that includes a node that is not present in the input graph.

\textbf{c) Search-based:} In these techniques, the goal is to find a counterfactual explanation by searching the counterfactual space. These counterfactual spaces can be exponentially based on the similarity methods that have been used to create them. Therefore, the challenge here is to design efficient search algorithms. One of the interesting works in this group is Global Counterfactual Explainer (GCFExplainer)~\cite{huang2023global}, which, unlike other counterfactual techniques, tries to find a set of counterfactual explanations for the whole input dataset or several subsets of the input space. First, they define the search space using the graph edit distance. The search space includes all the graphs from the same domain as input within a distant $\theta$. Then, vertex-reinforced random walk is used to find good counterfactual explanations. In~\cite{wellawatte2022model}, Molecular Model Agnostic Counterfactual Explanations (MMACE)
was proposed to find local counterfactual explanations by expanding the space around the source. Then, using Superfast Traversal, Optimization, Novelty, Exploration, and Discovery (STONED), the counterfactual explanations are detected, and by clustering and similarity check, the number of counterfactuals is reduced. Molecular Counterfactual Generator (MEG)~\cite{numeroso2021meg} also proposed a method to find the counterfactual explanations. They used RL to design an agent that searches the input space for a suitable counterfactual explanation. Related domain knowledge is necessary to ensure that the found counterfactual explanation is valid and follows the required domain constraints.

\subsubsection{Factual}
Unlike counterfactual methods, factual methods aim to find the subgraph or input features that are most influential on the model's prediction. Based on the literature, factual methods can be further classified into two groups based on how they integrate the explainability module. In self-interpretable models, explainability is part of the trained model, while in post-hoc techniques, explainability is built on top of the pre-trained model with fixed parameters.

\noindent\textbf{\textit{I}) Post-hoc:} As shown in Figure~\ref{fig:gnn_exp}, the post-hoc technique can be categorized into five categories based on their approach to provide explanations.

\textbf{a) Gradient-based:} In this approach, the gradient of the output with respect to the input features is used as a tool to measure the effect of the features on the prediction. One of the techniques used for explaining GNNs is Sensitivity Analysis (SA)~\cite{baldassarre2019explainability}, where they calculated the saliency score for the input using the squared norm of the gradients. In the same work, they used Guided Backpropagation (Guided-BP), which uses the same approach as SA with the slight change of clipping the negative gradients. In~\cite{pope2019explainability}, the authors borrowed several techniques used to explain CNN to provide explanations for GCN. Grad-CAM is their proposed technique, which falls into the gradient-based category. Grad-CAM is an extension of the CAM method, which relaxes the requirement for the one to the last layer to be convolutional. Since the gradient-based methods focus on explaining each prediction regarding their inputs, they can only provide instance-level explainability and cannot explain the target model as a whole.

\textbf{b) Decomposition-based:} These techniques consider the model output as a score that can be decomposed and back-propagated in a layer-wise manner to the input features. These methods analyze the model parameters to identify the relationships between input features and output predictions. A key aspect of these methods is that the sum of the decomposed terms equals the original prediction score. However, applying these methods to graphs is challenging because graphs include nodes, edges, and node features, making it difficult to distribute importance scores among edges, which hold essential structural information. In~\cite{pope2019explainability}, they also used CAM and Excitation-BP, which fall into the decomposition-based category. CAM is useful for models with a global average pooling layer before the last fully connected layer. In Excitation-BP, the final probability is the result of excitations from all the previous neurons and can be calculated through a backward pass. The excitations of all features of a node are combined to get their respective importance score. In Decomposition-based Explanation for Graph Neural Networks (DEGREE)~\cite{feng2023degree}, they aimed to provide a subgraph-level explanation by decomposing the feed-forward process of GNN. Given a target subgraph as the explanation, they considered part of the passed message: the target portion, which is the information from the target subgraph, and the background portion, which is the rest of the information. After calculating the score for each node through decomposition, they proposed an agglomeration-based algorithm to find the subgraph with the highest score. Another decomposition-based technique is GNN-LRP~\cite{schnake2021higher}, which, unlike the previous techniques, assigns a score to group edges (walks) to provide an explanation. The higher-order Taylor expansion of the target model function, along with the Layer-wised Relevance Propagation, is used to find suitable walks.

\textbf{c) Surrogate:} One of the main challenges in explaining the deep models is the complexity and non-linearity of the mapping from the input to the output. In surrogate methods, the idea is to train a simpler model on the neighboring samples of the target input. The challenge in using these methods for GNNs is the difficulty of selecting the neighboring samples. In~\cite{vu2020pgm}, authors proposed a Probabilistic Graphical Model-agnostic explainer for GNNs (PGM-Explainer) which trains a Bayesian network as the surrogate model to explain the prediction. It consists of three steps: data generation, variable selection, and structure learning. In the data generation step, the features of multiple nodes are randomly perturbed to create neighboring samples. Then, in the next step, the less important nodes are eliminated to simplify the generated data and speed up the explanation generation process. Finally, in the structure learning step, the Probabilistic Graphical Model, a Bayesian network, is trained using the Bayesian Information Criterion (BIC) and hill-climbing algorithm. GraphLime~\cite{huang2022graphlime} is an extension for the LIME algorithm to generate explanations for graph learning models. To create the local dataset for the surrogate model, it uses the N-hop neighbors of the target node and their prediction. Then, they used the Hilbert-Schmidt Independence Criterion (HSIC) Lasso to train the explainer and rank the features based on the learned weights of the model. In another work~\cite{duval2021graphsvx}, authors proposed Shapley Value Explanations for Graph Neural Networks (GraphSVX), using a mask generator to perturb the original dataset and build the local dataset for the surrogate model. Then, they fitted a Weighted Linear Regression (WLR) on the local dataset and used its weights as the explanation. A Model-Agnostic Relational Model Explainer (RelEx)~\cite{zhang2021relex} approach is more complex than previous methods since it uses a GCN as its surrogate model. RelEx first uses a BFS sampling technique to generate a local dataset of the target node, then queries the target model with the samples from the local dataset, and finally trains a GCN on the input-output pairs. While training GCN, they try to optimize a loss function, which finds a sparse mask as the explanation. In Distill n' Explain (DnX)~\cite{pereira2023distill}, they used SGCN as the surrogate model and trained it on the whole dataset. To distill the knowledge from the target model to the surrogate model, they adjusted the parameters of the surrogate model in a way that made both predictions the same. Then, to extract the explanation from the surrogate model, they used two techniques based on optimization and linear decomposition.

\textbf{d) Perturbation-based:} These techniques aim to find a small sub-graph and, in some cases, a small sub-feature as the explanation through perturbing the input until finding the suitable explanation based on a scoring function. One of the first attempts to explain GNNs was GNNExlpainer~\cite{ying2019gnnexplainer}, which gained a lot of attention and is among the most cited works in the field. They tried to find the subgraph and sub-feature explanation by maximizing the mutual information with the predicted label. They defined two masks for features and edges and learned them by optimizing a cross-entropy loss using gradient descent. Their technique is an instance-level explanation that can be expanded into multi-instance with minor changes and is suitable for node classification, graph classification, and link prediction tasks. Parameterized Explainer for Graph Neural Network (PGExplainer)~\cite{luo2020parameterized} is an extension to GNNExplainer and aims to provide a model level explanation for the various types of GNNs. They focus on finding an important subgraph by modeling the probability distribution of the edges and learning the parameters using an MLP. In another technique called GraphMask~\cite{schlichtkrull2020interpreting}, instead of finding the important edges, they tried to find edges that can be dropped while the model prediction is still the same. They replaced the dropped edges with a baseline learned vector to keep the graph structure unchanged. An MLP netwrok has been used to find the edges to drop. Zorro~\cite{funke2022z} proposes a new metric called RDT-Fidelity to measure the effectiveness of the GNN explanation. RDT-Fidelity is quantified by the expected validity score of an explanation over all possible perturbed inputs. Using a greedy approach, they find a sparse explanation that includes important nodes and features and maximizes the RDT-Fidelity. In a novel work by Wang et. el.~\cite{wang2021towards}, ReFine (pRE-training and Fine-tuning) is proposed, which aims to provide multi-grained explainability that includes both global and local explainability of the target GNN model. Their technique has two pre-training and fine-tuning steps to capture global and local explainability. SubgraphX~\cite{yuan2021explainability} is an instance-level explainability technique that uses Monte Carlo tree search to find an important subgraph. A scoring methodology based on Shapely values is used to rank the subgraphs and find the explanation. In Graph Structure-aware Explanation (GStarX)~\cite{zhang2022gstarx}, which follows the same approach as SubgraphX, they showed that the Shapley values are not suitable for GNN explainability since they are non-structure-aware and proposed to use structure-aware Hamiache and Navarro (HN) value to compute the importance score of the nodes to be included in the explanation. In~\cite{kubo2024xgexplainer}, the authors noted that using the same GNN model as the evaluator for extracted important subgraphs could fail to recognize out-of-distribution subgraphs. To address this, they proposed enhancing the explainer by training a dedicated GNN model specifically for evaluating the explainer’s output. More recently, Huang et al.~\cite{huang2024factorized} introduced K-FactExplainer to overcome the challenges faced by existing parameterized explainers, such as locality constraints and lossy aggregation, which limit their effectiveness in solving real-world problems.
    
\textbf{e) Generation-based:} In this group of techniques, generation-based techniques such as generative models or graph generators are used to extract the explanation. One of the famous works in this category is XGNN~\cite{yuan2020xgnn}, which uses an RL-based approach to generate the explanation subgraph, while checking for the validity of the explanation using several graph rules. Their approach aims to find a model-level explanation by generating a graph that maximizes the model prediction. Same as the XGNN, Reinforcement Learning Enhanced Explainer for Graph Neural Networks (RG-Explainer)~\cite{shan2021reinforcement} uses an RL-based approach to find an instance-level explanation in both node and graph classification tasks. This approach has three main components: starting point selection, iterative graph generation, and stopping criteria learning. The starting point is the most important node in a graph for the model prediction. After selecting the starting node, iteratively, new neighboring nodes are added by following the original input structure to ensure the generated explanation is valid. The reward for the RL agent is mutual information between the original label and the predicted label of the generated graph. Finally, a stopping criterion is learned to prevent generating very large graphs. GNNInterpreter~\cite{wang2022gnninterpreter} is a model-level explanation for any GNN architecture used for graph classification. Unlike XGNN, which uses domain knowledge to ensure the validity of the generated explanation, they used the knowledge learned by the target GNN itself. The similarity between the explanation graph embedding and the average embedding of all the graphs in the target class is the constraint for the explanation's validity. Another work is GFlowNets-based GNN Explainer (GFlowExplainer)~\cite{li2023dag}, which uses the generation property of Generative Flow Networks to overcome the shortcomings of the previous generation-based techniques. Their insight is to learn a generative policy that generates a distribution of connected subgraphs with probabilities proportional to their mutual information to provide an instance-level explanation for node and graph classification tasks. GEM~\cite{lin2021generative} leverages the principles of Granger causality to create ground-truth explanations for training the explainer. It measures the causal impact of each edge in the computational graph by comparing the model's loss with and without that edge. This distilled ground-truth information for the computational graph is then used to train a generative auto-encoder-based explainer. The explainer generates explanations for any instance by presenting a subgraph of the computational graph.

\noindent\textbf{\textit{II}) Self-interpretable:} In self-interpretable models, unlike post-hoc techniques, the explainability is part of the model itself, and the explanation is found at the same time as the prediction. The subgraph extraction module uses information or structural constraints to find the important subgraph.

\textbf{a) Information Constraint:} In the information constraint-based techniques, instead of limiting the size of the explanation, the information bottleneck principle is used to limit the information in the subgraph. Graph Information Bottleneck (GIB)~\cite{yu2020graph} is one of the first works that uses information bottleneck to recognize the important subgraph. They used a bi-level optimization to maximize GIB. Variational GIB (VGIB)~\cite{yu2022improving} uses the same approach with a different compression technique to make the bi-level optimization more efficient. Miao et al.~\cite{miao2022interpretable} proposed Graph Stochastic Attention (GSAT), which uses an attention mechanism to inject stochasticity to block the information from the task-irrelevant graph to find the interpretation. The same authors later proposed Learnable Randomness Injection (LRI)~\cite{miao2022interpretableLRI}, which uses the same concept for extracting the explainability.

\textbf{b) Structural Constraint:} In another group of self-interpretable GNN techniques, the goal was to impose structural constraints to find the important subgraph. Self-Explainable GNN (SE-GNN)~\cite{dai2021towards}, as one of the early works in this category, uses the k-nearest labeled nodes to provide explainability for the unlabeled node. The k-nearest labeled nodes are selected based on the node similarity and the local structure similarity. Based on Discovering Invariant Rationales (DIR), Wu et al.~\cite{wu2022discovering} proposed a technique that learns to split the input graph into causal and non-causal subgraphs. Two classifiers are trained on the causal and non-causal parts to calculate the joint prediction and extract the explanation. Prototype Graph Neural Network (ProtGNN)~\cite{zhang2022protgnn} uses prototype learning, which classifies samples by checking the similarity between the input and several learned prototypes. The subgraphs with high similarity scores can be used as the explanation. Interpretable Graph Neural Networks with Graph Kernels (KerGNN)~\cite{feng2022kergnns} integrate the graph kernels into the message-passing calculation of GNNs and use the trained graph filters to reveal the local graph structure in order to improve the model's interpretability.

\begin{figure}[H]
    \centering
    \includegraphics[width=0.7\linewidth]{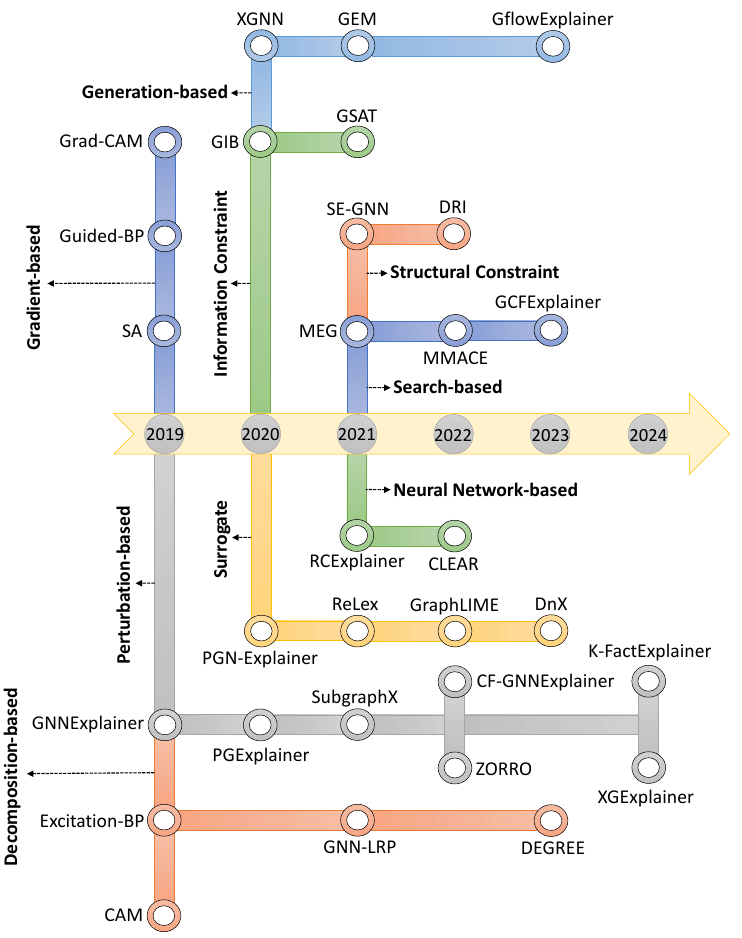}
    \caption{Timeline of GNN explainers, representing most significant exlpainers of each category.}
    \label{fig:gnn_exp_time}
\end{figure}

Figure~\ref{fig:gnn_exp_time} shows the timeline for the GNN explainers since 2019. To make the provided timeline more concise, some of the less significant works are dismissed.

Table~\ref{tab:gnn_exp} summarizes all the GNN explainers. Following the taxonomy proposed in Figure~\ref{fig:gnn_exp}, in the ``Category" column, we have Factual Self-interpretable, Factual Post-hoc, and Counterfactual Post-hoc. For the downstream task that the explainer can be used for there are three tasks of graph classification, node classification, and node ranking. Different GNN explainers either directly generate the important subgraph or let the user generate the subgraph by providing them with the importance score or weight for nodes, edges, or features. This classification is shown in the ``Output" column.

As explained in the provided taxonomy (Figure~\ref{fig:gnn_exp}) the explainability of GNNs can be post-hoc and intrinsic. Since, in most cases of explainers for malware detection, the goal is to provide explainability for a pre-trained model and training a new self-interpretable GNN might be infeasible, the focus should be on the post-hoc techniques. Between factual and counterfactual approaches, as the goal of the malware detection task is to find the most representative subgraph for the made decision, factual techniques seem more appropriate. At the same time, while all the currently available databases and approaches focus on classifying a whole graph as malware or benign and no labels are provided for each node, only techniques with graph classification as their downstream task can be useful for malware detection. Based on the explained criteria, we tried to label different techniques as recommended or not recommended for malware detection in the last column in Table~\ref{tab:gnn_exp}.

\begin{table}
\centering
\caption{Summary of GNN explainers (FS: Factual Self-interpretable, FP: Factual Post-hoc, CP: Counterfactual Post-hoc, NdR: Node Ranking, GC: Graph Classification, NC: Node Classification, R: Recommended, and NR: Not Recommended).}
\label{tab:gnn_exp}
    \begin{tabular}{p{3.2cm} p{1.2cm}<{\centering} p{3.5cm}<{\centering} p{1.8cm}<{\centering} p{2.2cm}<{\centering} p{1.3cm}<{\centering}} 
        \toprule
        \rule{0pt}{0.3cm} 
        \textbf{Technique} & \textbf{Category} & \textbf{Subcategory} & \textbf{Task} & \textbf{Output} & \textbf{Malware Detection} \\
        \hline
        GIB~\cite{yu2020graph} & FS & Information Constraint & GC & Subgraph & NR \\
        
        VGIB~\cite{yu2022improving} & FS & Information Constraint & GC & Subgraph & NR \\
        
        GSAT~\cite{miao2022interpretable} & FS & Information Constraint & GC & Subgraph & NR \\
        
        LRI~\cite{miao2022interpretableLRI} & FS & Information Constraint & GC & Node & NR \\
        
        DIR~\cite{wu2022discovering} & FS & Structural Constraint & GC & Subgraph & NR \\
        
        ProtGNN~\cite{zhang2022protgnn} & FS & Structural Constraint & GC & Subgraph & NR \\
        
        SEGNN~\cite{dai2021towards} & FS & Structural Constraint & NC & Node, Edge & NR \\
        
        KER-GNN~\cite{feng2022kergnns} & FS & Structural Constraint & GC, NC & Subgraph & NR \\
        
        CAM~\cite{pope2019explainability} & FP & Decomposition-based & GC, NC & Node & R \\
        
        Excitation-BP~\cite{pope2019explainability} & FP & Decomposition-based & GC, NC & Node & R \\
        
        DEGREE~\cite{feng2023degree} & FP & Decomposition-based & GC, NC & Node & R \\
        
        GNN-LRP~\cite{schnake2021higher} & FP & Decomposition-based & GC, NC & Node & R \\
        
        SA~\cite{baldassarre2019explainability} & FP & Gradient-based & GC, NC & Node & R \\
        
        Guided-BP~\cite{baldassarre2019explainability} & FP & Gradient-based & GC, NC & Node & R \\
        
        Grad-CAM~\cite{pope2019explainability} & FP & Gradient-based & GC, NC & Node & R \\
        
        PGM-Ex~\cite{vu2020pgm} & FP & Surrogate & GC, NC & Edge & R \\
        
        GraphLime~\cite{huang2022graphlime} & FP & Surrogate & NC & Node, Feature & NR \\
        
        GraphSVX~\cite{duval2021graphsvx} & FP & Surrogate & GC, NC & Node, Feature & R \\
        
        ReLex~\cite{zhang2021relex} & FP & Surrogate & NC & Node, Edge & NR \\
        
        DnX~\cite{pereira2023distill} & FP & Surrogate & NC & Node & NR \\
        
        GraphMask~\cite{schlichtkrull2020interpreting} & FP & Perturbation-based & GC & Edge & R \\
        
        GNNExplainer~\cite{ying2019gnnexplainer} & FP & Perturbation-based & GC, NC & Edge, Feature & R \\
        
        PGExplainer~\cite{luo2020parameterized} & FP & Perturbation-based & GC, NC & Edge & R \\
        
        ReFine~\cite{wang2021towards} & FP & Perturbation-based & GC & Edge & R \\
        
        ZORRO~\cite{funke2022z} & FP & Perturbation-based & NC & Node & NR \\
        
        SubgraphX~\cite{yuan2021explainability} & FP & Perturbation-based & GC, NC & Subgraph & R \\
        
        GstarX~\cite{zhang2022gstarx} & FP & Perturbation-based & GC, NC & Node & R \\

        XGExplainer~\cite{kubo2024xgexplainer} & FP & Perturbation-based & GC, NC & Node, Edge & R \\

        K-FactExplainer~\cite{huang2024factorized} & FP & Perturbation-based & GC, NC, & Edge & R \\
        
        XGNN~\cite{yuan2020xgnn} & FP & Generation-based & GC & Subgraph & R \\
        
        RGExplainer~\cite{shan2021reinforcement} & FP & Generation-based & GC, NC & Subgraph & R \\
        
        GNNInterpreter~\cite{wang2022gnninterpreter} & FP & Generation-based & GC & Subgraph & R \\
        
        GflowExplainer~\cite{li2023dag} & FP & Generation-based & GC, NC & Subgraph & R \\
        
        GEM~\cite{lin2021generative} & FP & Generation-based & GC, NC & Subgraph & R \\
        
        MMACE~\cite{wellawatte2022model} & CP & Search-based & GC, NC & Subgraph & NR \\
        
        MEG~\cite{numeroso2021meg} & CP & Search-based & GC, NC & Subgraph & NR \\
        
        GCFExplainer~\cite{huang2023global} & CP & Search-based & GC & Subgraph & NR \\
        
        RCExplainer~\cite{bajaj2021robust} & CP & Neural Network-based & GC, NC & Edge & NR \\
        
        CLEAR~\cite{ma2022clear} & CP & Neural Network-based & GC & Edge & NR \\
        
        GREASE~\cite{chen2022grease} & CP & Perturbation-based & NdR & Edge & NR \\
        
        CF\textsuperscript{2}~\cite{tan2022learning} & CP & Perturbation-based & GC, NC & Edge & NR \\
        
        CF-GNNExplainer~\cite{lucic2022cf} & CP & Perturbation-based & NC & Edge & NR \\
        \bottomrule
    \end{tabular}
\end{table}

\subsection{GNN Explainability Evaluation Metrics}
Since the explainability of AI models is a new concept, evaluation metrics that are suitable for this specific task need to be provided. Several metrics were introduced in the literature to measure the goodness of the explanations and compare the GNN explainers. In the following, we bring a list of the most commonly used metrics with a brief explanation:

\begin{itemize}[leftmargin=*]
    \item \textbf{Fidelity:} This metric measures how faithful the explanation is to the model's prediction. It can be formulated in two forms: Fidelity\textsuperscript{+} and Fidelity\textsuperscript{-}. The difference between the prediction of the original sample and the explanation is Fidelity\textsuperscript{-}. On the other hand, the difference between the prediction of the original sample and an input, in which the explanation is removed from the original sample, is Fidelity\textsuperscript{+}. The formula for calculating Fidelity\textsuperscript{-} and Fidelity\textsuperscript{+} are as follows:

    \begin{equation}
        Fidelity^{+} = \frac{1}{N}\sum_{i=1}^{N}\left ( f\left ( G_i \right ) - f\left ( G_i-G_{s_i} \right ) \right ),
    \end{equation}

    \begin{equation}
        Fidelity^{-} = \frac{1}{N}\sum_{i=1}^{N}\left ( f\left ( G_i \right ) - f\left ( G_{s_i} \right ) \right ),
    \end{equation}

    where $G_i$ is the $i$th original sample, $G_{s_i}$ is the important subgraph corresponding to $i$th original graph, $N$ is the number of samples, and $f$ is the GNN model.
    
    \item \textbf{Sparsity:} A good explanation should include only the most important nodes and edges, while discarding the ones that are irrelevant. Therefore, a more sparse explanation is the better one. Sparsity can be measured by one minus the number of nodes or edges in the explanation over all the nodes or edges in the original graph.

    \begin{equation}
        Sparsity = 1-\frac{1}{N}\sum_{i=1}^{N}\frac{\left | G_{s_{i}} \right |}{\left | G_{i} \right |}.
    \end{equation}

    In the above formula, operator $| . |$ returns the number of edges or nodes.
    
    \item \textbf{Validity:} The goal of explainability is to find the most important subgraph for the GNN's prediction; therefore, the prediction of the explanation as the input should be the same as the original input prediction. Validity is the proportion of the explanation that generated the correct label.
    
    \item \textbf{Efficiency (Time):} Another useful evaluation metric is efficiency, which measures the average time cost for generating the explanation for the samples in the dataset. 
    
    \item \textbf{Accuracy:} One commonly used metric when evaluating explainers on synthetic datasets is accuracy. Since ground truth explainability is available in the synthetic datasets, the accuracy of the explainer can be measured.
\end{itemize}

\subsection{Explainable Models for Malware Detection}
One of the first works that tried to provide some explainability for malware detection was done by Arp et al.~\cite{arp2014drebin}. In their paper, they proposed an SVM-based detection model for Android malware, which provides some textual explanations of the made decisions. They did a broad static analysis of the malware files in the feature extraction step. They extracted several features from eight different categories, namely hardware components, requested permissions, App components, filtered intents, restricted API calls, used permissions, suspicious API calls, and network addresses. Then, the extracted features were embedded into a vector space, where the dimension is the same as the number of all existing features. After training the SVM model, the top k features with the highest weights are stored, and some pre-designed sentence templates for each feature group are used to provide the textual explanation.

In~\cite{CFG_20}, they proposed a context-aware, adaptive, and scalable Android malware detector named CASANDRA. First, through static analysis, they extracted the Contextual API Dependency Graph (CADG) as input features to the framework. Then, the security-sensitive parts of the CADG were extracted, and a confidence-weighted learner was trained for malware detection. For explainability, the top features with the highest weights are selected and reported.

Bose et al.~\cite{bose2020explaining} proposed a framework for explaining deep neural network test results by interpolating between different samples in different layers. They incorporated their framework to analyze MalConv~\cite{raff2018malware}, a CNN-based model for malware detection on Windows executable files. Their framework uses three techniques to analyze the target model's results, including gradient analysis, interpolation between samples, and filter correlation. Although they used their framework for a CNN model for malware detection, it can be used to analyze any neural network performing classification tasks.

In~\cite{yan2021effective}, the authors proposed a rule extraction method for deep neural network-based malware detection based on network traffic analysis features. They generated two decision trees, namely the hidden-output tree and the input-hidden tree. Since the output of one tree is the input of the other tree, they can easily merge these two and get the final rule tree. They compared their method performance with several well-known ML techniques such as Bagging, AdaBoost, kNN, and RandomForest and other state-of-the-art works. While their method had comparable performance to other techniques, it provided an acceptable level of explainability about the decisions made.

Kinkead et al.~\cite {kinkead2021towards} proposed an explainability method for a CNN-based malware detection model and compared their result with LIME, showing that their method is effective in terms of both detection and explainability. They used CNN with the same architecture as in~\cite{mclaughlin2017deep} and the DREBIN android malware dataset for their experimental analysis. Their goal was to extract the activation of the opcode sequence from the CNN model and compare it with the activations calculated using LIME. Their result showed that, while their model has comparable performance to other state-of-the-art techniques, it highlights the similar areas in the opcode sequence as LIME and gives more weight to the same opcodes in its decision.

In~\cite{wu2021android}, Wu et al. combined multi-layer perception with an attention mechanism and used it as a classifier for malware detection. The input features for their model were API calls and permission used by the Android malware. Upon detection, they used the attention mechanism weights to rank the features and selected the top $n$ features to generate the textual explanation. They generated the final textual explanation by building a semantic database that maps each feature to a description based on an Android developer documentation.

CFGExlpainer~\cite{CFG_4} is the first attempt to propose a technique for explaining GNN-based malware detection frameworks that use CFGs. CFGExlpainer ranks nodes based on their importance for the classification task and, at the same time, generates important subgraphs constructed using the top-ranked nodes. To rank the nodes based in their importance, they used two feed-forward neural networks including the node embedding generated by the target GNN and the GNN classification output. After ranking the nodes, the explanation subgraphs are generated by iteratively removing the least important nodes.

Pan et al.~\cite{pan2022hardware} proposed a malware detection method using hardware-based features such as hardware performance counter (HPC) and embedded trace buffer (ETB). After performing data accusation, they used a decision tree and LSTM as their detection model. As mentioned before, decision trees are self-explanatory, and a tree traversal technique is used for the decision explanation. On the other hand, LSTM models do not have explanatory properties, and an extra step is required to explain their decision. They used a linear regression-based approach to find the most significant features of the trained model.

In another work that focused on the explainability of the GNN-based malware detection model, Wramsley et al.~\cite{warmsley2022survey} compared five different explainer techniques, including GNNExplainer, PGExplainer, DeepLIFT, Grad-CAM, and CFGExplainer. They evaluated these methods using Fidelity, Sparsity, and Harmonic Fidelity as their metrics. Their target model was GIN and GCN trained on the extracted CFGs from binary files of malware families.

In~\cite{ullah2022explainable,hsupeng2022explainable,alani2022paired}, authors used LIME and SHAP to add explainability to their proposed methods for malware detection. Ullah et al.~\cite{ullah2022explainable} proposed a technique that combines textual and visual features extracted from network flow to detect Android malware. They used a BERT-based method for the textual features and a CNN model for the visual representation extracted from the network data. They incorporated both LIME and SHAP to provide explainability. In~\cite{hsupeng2022explainable}, they also used network data as their model input and used XGBoost and SHAP for detection and explanation. Alani et al.~\cite{alani2022paired} used recursive feature elimination based on feature importance to minimize the number of features used in their Android malware detection technique. They aimed to achieve a lightweight detection model by minimizing the number of used features. For explainability, they used SHAP and showed the top 10 features of the dataset.

In~\cite{he2022msdroid}, authored proposed a framework for detecting Android malware using call graph and GNN-based classifier. They also provided three types of explainability in their framework: Suspicious API Usage, Calling Heat Graph, and Similar Behavioral Snippets. 

In~\cite{alani2023xmal}, Alani et al. also used the same approach as~\cite{alani2022paired}, combining recursive feature elimination and the SHAP technique in order to propose a lightweight memory-based explainable obfuscated-malware detector. Authors of~\cite{ACG_11} extracted the FCGs of APK files and used the API calls belonging to Android critical packages as inputs for their detection model input. Then, they used a 1D-CNN for classification and decision making and SHAP for explaining the predictions.

Authors in \cite{mohammadian2024explainablemalwaredetectionintegrated} explored the integration of graph reduction techniques with GNNs to enhance malware detection efficiency and interpretability. They leveraged CFGs and FCGs to represent program execution and applied GNNExplainer to improve transparency in GNN decision-making. Their findings indicate that the proposed graph reduction methods significantly reduce graph size while maintaining high detection performance, striking a balance between computational efficiency and model explainability.

One of the most recent works that makes use of an explainer along with a malware detection framework is~\cite{FCG_15}, in which the importance of each node in the FCG is calculated. The node embedding vector is multiplied by the weight vector of the output layer in order to get the importance vector of the FCG nodes.

To conclude this section, the summary of the malware detection approaches is provided in Table~\ref{tab:mal_exp}. By reviewing the works presented in Table~\ref{tab:mal_exp}, it is clear that there is a lack of work on explanation of malware detection techniques that use state-of-the-art GNN models. With the recent interest in using different call graphs as the input feature for the malware detection task, more attention to the explainability of GNNs is required for the malware detection task.

\begin{table*}
    \setlength{\tabcolsep}{4pt}
    \centering
    \caption{Explainable malware detection approaches.}
    \resizebox{\textwidth}{!}{%
    \begin{tabular}{lccc}
        \toprule
        \textbf{Reference} & \textbf{Target} & \textbf{ML/DL Algorithm} & \multicolumn{1}{c}{\textbf{Explainability Technique}} \\
        \midrule
        \cite{arp2014drebin} & Android & SVM & Feature Relevance / Textual Explanation \\
        
        \cite{CFG_20} & Android & CW Learner & Feature Ranking by Weights \\
        
        \cite{bose2020explaining} & Windows & CNN & Gradient Analysis / Samples Interpolation / Filter Correlation \\
        
        \cite{yan2021effective} & Android & MLP & Rule Extraction \\
        
        \cite{kinkead2021towards} & Android & CNN & Opcode Sequence Activation \\
        
        \cite{wu2021android} & Android & MLP + AM & Textual Explanation \\
        
        \cite{CFG_4} & Windows & GCN & Important Nodes and Subgraph \\
        
        \cite{pan2022hardware} & IoT & DT / LSTM & Linear Regression \\
        
        \cite{warmsley2022survey} & Windows & GIN / GCN & GNNExplainer / PGExplainer / DeepLIFT / Grad-CAM / CFGExplainer \\
        
        \cite{ullah2022explainable} & Android & BERT + CNN & LIME / SHAP \\
        
        \cite{hsupeng2022explainable} & Android & XGBoost & SHAP \\
        
        \cite{alani2022paired} & Android & RF & SHAP \\
        
        \cite{he2022msdroid} & Android & GCN & Suspicious API Usage / Calling Heat Graph / Similar Behavioral Snippets \\
        
        \cite{alani2023xmal} & Windows & XGBoost & SHAP \\
        
        \cite{ACG_11} & Android & CNN & SHAP \\
        
        \cite{FCG_15} & Android & GraphSAGE & Node Importance of FCG \\
        
        \cite{mohammadian2024explainablemalwaredetectionintegrated} & Windows & GCN & GNNExplainer \\
        \bottomrule
    \end{tabular}
    }
    \label{tab:mal_exp}
\end{table*}

\subsection{Challenges and Future Directions}
In this section, we aim to conclude our study on Explainable malware detection approaches by briefly reviewing challenges in the realm of XAI and the future directions toward explainability of intelligent malware detection systems.

The relationship between explainability and performance has long been debated in XAI. Traditional views suggest a tradeoff, where increasing model complexity often improves performance but reduces explainability. However, Rudin~\cite{rudin2019stop} provides evidence that simpler models, such as linear regressors and rule-based systems, can achieve performance comparable to more complex models like neural networks and random forests, particularly in well-defined industrial settings. While complex models are often preferred for their flexibility and ability to capture intricate patterns in environments with abundant data, this very complexity—along with their deep architectures—often leads to reduced explainability.

Dziugaite et al.~\cite{dziugaite2020enforcing} focused on reducing the explainability-performance tradeoff by enforcing explainability constraints during model training. Their research suggests that it is possible to design both performant and interpretable models, challenging the assumption that a tradeoff must exist and highlighting a trend that might be reversible. Future XAI techniques should aim to optimize this balance to cater to specific operational needs and user preferences. Emerging research indicates that the perceived inevitable tradeoff between explainability and performance in XAI can be mitigated. With a growing number of studies challenging this notion, there is optimistic potential for developing methods that retain or even enhance model understandability without compromising on performance, potentially leading to broader acceptance and usability of AI systems across various domains.

Another challenge evident in the literature is the requirement to provide a clear definition of explainability. This concept is essential for developing a framework that allows the community to introduce and refine methods and techniques effectively. A definition of explainability can be proposed as the capacity of a model to clarify its functions to an audience. It highlights the initial steps toward establishing a common understanding that can support meaningful discussions and further developments in the field. While the proposed concept provides a starting point, it also calls for an agreed-upon metric or set of metrics that can quantitatively assess how well a model meets this definition of explainability.

Current efforts in the field include various metrics to measure the effectiveness of XAI, such as the goodness checklist, explanation satisfaction scale, and others that assess the impact of explanations on user trust and model reliability. However, there is still a significant gap in the availability of general, quantifiable metrics that can universally evaluate XAI techniques. More comprehensive and standardized measurement tools are needed to strengthen the foundation for comparing and improving XAI methods.

Another challenge in this area is proposing effective methods for explaining complex deep learning models. The research community has not agreed on the terminology or the categorization of XAI techniques, as evidenced by the interchangeable usage of terms such as feature relevance and feature importance and the lack of standardized definitions for post-modeling visualization methods like saliency maps and heatmaps. Moreover, the complexity inherent in nonlinear deep learning models, which contrasts with the straightforward interpretability of linear models, presents significant challenges in interpretation, especially in high-dimensional spaces, despite various innovative visualization techniques that attempt to clarify the internal mechanisms of these models.

The challenges discussed in XAI are particularly pronounced in the context of malware detection, where conventional explainability approaches may not seamlessly translate to the unique demands of this task. Traditional explainability metrics, such as fidelity and comprehensibility, often assume structured, interpretable feature spaces, which may not be suitable for malware detection models that operate on complex, obfuscated, and high-dimensional feature representations. Malware detection often relies on intricate behavioral patterns making it difficult to generate explanations that are both meaningful and faithful to the model’s decision-making process. Furthermore, in GNN-based malware detection, the sheer size and structural complexity of graph representations—such as CFGs and FCGs—pose additional interpretability challenges. Large graphs make it computationally expensive to generate explanations, while the non-Euclidean nature of graph data limits the effectiveness of traditional feature-based XAI methods. Addressing these challenges requires the development of domain-specific XAI techniques and evaluation frameworks that align well with the realities of the malware detection task, ensuring both interpretability and reliability without compromising the detection performance.

\section{Conclusion}
This survey has presented a thorough exploration of recent advances in malware detection, emphasizing the transformative potential of graph learning and explainability. By dissecting the critical components required for malware detection including malware analysis, dataset collection, feature engineering, graph reduction, graph embedding, and learning we highlighted the interconnected nature of these processes and their collective contribution for effective detection of malware. The review of explainability, particularly those techniques tailored for GNNs, underscores the importance of transparency and interpretability in ensuring the trustworthiness of AI-driven solutions.

Our analysis revealed significant strides in leveraging graph-based techniques to model and analyze malware behavior. Graph reduction and embedding methods have addressed challenges related to scalability and efficiency, while explainability has bridged the gap between high detection accuracy and actionable insights. Despite these advancements, challenges remain, including the need for more robust data, scalable reduction techniques for extremely large graphs, and methods that can explain increasingly complex models and scenarios.

Future research should focus on integrating these components into cohesive frameworks that address the dynamic and evolving nature of malware threats. Specific areas for advancement include the development of standardized benchmarks for evaluating graph-based methods for malware detection, enhanced graph explainer for real-time applications, and exploration of novel embedding techniques that capture both static and dynamic malware behaviors.

By addressing these challenges, the field can advance toward building more robust, scalable, and interpretable malware detection systems. This study serves as a foundation for future research, offering a roadmap to navigate the complexities of applying graph learning and explainability in this critical domain of cybersecurity.

\textbf{Funding Declaration} There is no funding for this research.


\bibliography{sn-bibliography}

\end{document}